\documentclass[preprint]{jpsj2}
\usepackage{amsmath}
\usepackage[dvips]{graphicx,color}
\makeatletter

\@addtoreset{equation}{section}
\makeatother

\def\runauthor{Online-Journal Subcommittee}
\title{
Crystalline-Electric-Field Effect on the Resistivity \\of \\Ce-based Heavy Fermion Systems 
}
\author{%
Yasutaka \textsc{Nishida}\thanks{E-mail address: 
y-nishida@blade.mp.es.osaka-u.ac.jp}, Atsushi \textsc{Tsuruta} 
and Kazumasa \textsc{Miyake}
}
\inst{%
Department of Materials Engineering Science,\\
Graduate School of Engineering Science, Osaka University, \\
Toyonaka, Osaka 560-8531
}
\recdate{\today}
\abst{%
The behavior of the resistivity of Ce-based heavy fermion systems is studied using
a 1/$N$-expansion method \textit{\'{a}} $la$ Nagoya, where $N$ is the spin-orbital 
degeneracy of f-electrons. The 1/$N$-expansion is performed in terms of the auxiliary 
particles, and a strict requirement of the local constraints is fulfilled for each order 
of $1/N$. The physical quantities can be calculated over the entire temperature range by 
solving the coupled Dyson equations for the Green functions self-consistently at each 
temperature. This $1/N$-expansion method is known to provide asymptotically exact results 
for the behavior of physical quantities in both low- and high-energy regions when it is 
applied to a single orbital periodic Anderson model (PAM). On the basis of a generalized PAM 
including crystalline-electric-field splitting with a single conduction band, the pressure 
dependence of the resistivity $\rho$ is calculated by parameterizing the effect of pressure 
as the variation of the hybridization parameter between the conduction electrons and f-electrons.
The main result of the present study is that the double-peak structure of the $T$-dependence of 
$\rho(T)$ is shown to merge into a single-peak structure with increasing pressure.
}
\kword{%
Ce-based heavy fermions, electrical resistivity, 1/$N$-expansion method, crystalline electric field
}
\begin{document}
\maketitle
\section{Introduction}
The resistivity of heavy fermions such as Ce-based compounds (Kondo-lattice systems)
changes markedly with decreasing temperature. In the high-temperature region
the conduction electrons are decoupled from the f-electrons, which behave as localized
moments, exhibiting the Kondo effect that results in the logarithmic temperature dependence
of the resistivity. On the other hand, below the characteristic temperature $E_0$
(corresponding to the Kondo temperature in an impurity model) conduction electrons
are scattered coherently by the f-electrons, and a Fermi-liquid state with quasi-particles of increased
mass is formed. 
The detailed $T$ dependence of the resistivity $\rho(T)$ around $T=E_0$ can be roughly 
classified into two cases. In the first case, $\rho(T)$ has a single-peak structure 
as observed in $\textrm{CeCu}_6$\cite{Sumi}. 
In the second case, $\rho(T)$ has a double-peak structure arising from the 
crystalline-electric-field (CEF) effect, as observed in $\textrm{CeAl}_2$\cite{Onu},
$\textrm{CeCu}_2$\cite{Vargoz}, $\textrm{CeAu}_2\textrm{Si}_2$\cite{Link}
$\textrm{CeCu}_2\textrm{Si}_2$\cite{Holmes}, and $\textrm{CeCu}_2\textrm{Ge}_2$\cite{Jac}, for example.
Experimental studies of the effect of pressure on Ce-based compounds have revealed the fact that
the double peak in the resistivity curve
arising from the CEF effect tends to merge into a single peak as the pressure is increased.
In the 1980's the effect of CEF on the impurity Kondo effect was investigated theoretically by
various methods\cite{mae,kawa2,yama,han}.
However, to our knowledge, the effect of CEF in heavy fermions (lattice systems) has 
not been adequately investigated to date. In this work, we study the influence of
the CEF effect on the $T$-dependence of $\rho$ of Ce-based heavy fermions, 
and clarify the origin of the double-peak structure of $\rho(T)$.

To this end, we investigate an infinite-$U$ generalized periodic Anderson model (PAM), 
including the CEF splitting of f-electrons,
by using a 1/$N$-expansion method
\textit{\'{a}} $la$ Nagoya, where $N$ is the spin-orbital degeneracy of correlated electrons
\cite{Ono1,Ono2}. This 1/$N$-expansion method is performed in terms of the auxiliary
particles together with strict requirement of the local constraints in each order of 1/$N$ so as
to rigorously take into account the effects of the infiniteness of $U$.
As a result, the theory is free from an artificial transition of slave-boson condensation 
and gives correct behaviors of physical quantities in both low- and high-energy regions. 
Applying this method, we discuss the pressure effect of $\rho(T)$, and show that the 
result is consistent with the tendency observed in Ce-based heavy fermion systems.

The paper is organized as follows. In $\S 2$, we introduce the model Hamiltonian, a generalized
PAM, and the concept of the $1/N$-expansion method \textit{\'{a}} $la$ Nagoya.
In $\S 3$, we show the result at $T=0$ (the case of absolute zero temperature). In $\S 4$,
we show results at a finite temperature by using a self-consistent $1/N$-expansion beyond
the low-temperature approximation. Finally, we show the effect of pressure on the resistivity 
$\rho(T)$ in $\S 5$. Three appendices are given detailing
the calculation and algorithm of the $1/N$-expansion method.
\section{Model and Formal Preliminaries}
\subsection{Infinite-$U$ generalized periodic Anderson model}
We start with an infinite-$U$ periodic Anderson model (PAM) with an f-electron in a manifold of $J=5/2$\cite{yama1}, which is a simple and realistic model for 
the Ce-based heavy fermion systems: Our model Hamiltonian is given by
%
%
\begin{align}
H= &H_c+H_f+H_{cf},\\
H_c= & \sum_{{\bf k}\sigma} \varepsilon_{{\bf k}\sigma}c_{{\bf k}\sigma}^{\dag}c_{{\bf k}\sigma},\\
H_f= &\sum_{i\Gamma}E_{i\Gamma}f_{i\Gamma}^{\dag}f_{i\Gamma},\\
H_{cf}= &\frac{1}{\sqrt{N_L}}\sum_{{\bf k},i,\Gamma,\sigma}
(V_{{\bf k}\Gamma \sigma}\textrm{e}^{-\textrm{i}{\bf k}\cdot{\bf R}_i} c_{{\bf k}\sigma}^{\dag}f_{i\Gamma}b_{i}^{\dag}+\textrm{h.c.}),
\end{align}
%
%
\begin{figure}
\begin{center}
\includegraphics[scale=0.45]{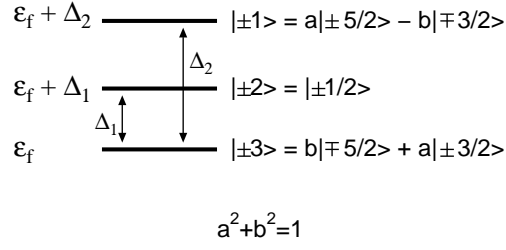}
\end{center}
\caption{CEF level scheme of $\textrm{Ce}^{3+}$ ion in tetragonal symmetry.}
\label{fig1}
\end{figure}
where $c_{{\bf k}\sigma}^{\dag}$ is the creation operator for the conduction electron
with wave vector ${\bf k}$ and spin $\sigma$, and $\varepsilon_{{\bf k}\sigma}$
is the energy of the conduction electron with wave vector $\bf k$ and spin $\sigma$. 
$N_L$ denotes the number of the lattice sites, and $i$ stands for the site index.
Here, we introduce the slave-boson, which represents the $f^0$-state at each 
lattice point, following Coleman \cite{Col}:
$b_i^{\dag}$ is the creation operator for the slave-boson at the $i$-th site.
$f_{i\Gamma}^{\dag}$ is the creation operator for the
pseudo-fermion which represents the $f^1$-state with
the CEF level $| \Gamma \rangle = | \pm 1\rangle ,| \pm 2\rangle,| \pm 3\rangle$.
$V_{{\bf k}\Gamma \sigma}$ stands for the mixing between the conduction electron 
with $\bf{k}$, $\sigma$ and the f-electron with the CEF state of $\Gamma$.
We discuss the property of the generalized PAM introduced above assuming
under a tetragonal symmetry of the crystal field.
Following Kontani $et\ al.$\cite{kon}, the mixing $V_{{\bf{k}}\Gamma\sigma}$ is expressed as
%
%
\begin{align}
V_{{\bf k}\Gamma \sigma}&=\sum_{M}O_{\Gamma M}V_{{\bf k}M\sigma},\\
V_{{\bf k}M\sigma}&=V_0{\sqrt{\frac{4\pi}{3}}} \Bigl\{ -2\sigma \sqrt{\frac{(\frac{7}{2}-2M\sigma)}{7}} Y_{l=3}^{M-\sigma}(\Omega_{{\bf k}})  \Bigr\},
\end{align}
where $M$ is the z-component of the angular momentum $J$ of an f-electron: $M=J_z\ (J=5/2)$,
and $Y_{l}^{m}(\Omega_{ k})$ is the spherical harmonic function.
In the tetragonal symmetry, we can parameterize each CEF-level as follows: (See Fig. \ref{fig1}.)
%
%
\begin{align}
E_{i|\pm 1\rangle}=\varepsilon_{f}+\Delta_2,\ 
E_{i|\pm 2\rangle}=\varepsilon_{f}+\Delta_1,\  
E_{i|\pm 3\rangle}=\varepsilon_{f}.
\end{align}
$\varepsilon_f$ is measured from the chemical potential $\varepsilon_f \equiv \varepsilon^{(0)}_f-\mu$ and 
$\varepsilon_f^{(0)}$ is defined as the lowest atomic level of f-electrons measured from the center of the conduction electron band.
The coefficient $O_{\Gamma M}$ in eq. (2.5) is a 6 $\times$ 6 orthogonal matrix, which is given as follows:
%
%
\begin{align}
O_{\Gamma M}=
 \left(
\begin{array}{c|cccccc}
  \Gamma  \setminus M  & 5/2 & 3/2 & 1/2 & -1/2 & -3/2 & -5/2 \\ \hline
    +3& 0 & a & 0 & 0 & 0 & \sqrt{1-a^2} \\
    +2&0 & 0 & 1 & 0 & 0 & 0 \\
    +1&a & 0 & 0 & 0 & -\sqrt{1-a^2} & 0 \\
    -1&0 & -\sqrt{1-a^2} & 0 & 0 & 0 & a \\
    -2&0 & 0 & 0 & 1 & 0 & 0 \\
    -3&\sqrt{1-a^2} & 0 & 0 & 0 & a & 0 \\
\end{array}
 \right)_.
\end{align}
Hereafter, all the energies are measured from the chemical potential $\mu \equiv 0$.

Next, we recapitulate the $1/N$-expansion method \textit{\'{a}} $la$ Nagoya. 
It is straightforward to extend the work by \=Ono $et\ al.$ for a single-band PAM as detailed in ref. \citen{Ono2}. 
In order to ensure equivalence between the present model Hamiltonian eqs. (2.1)$\sim$(2.4) and 
the original infinite-$U$ generalized PAM, we must treat the problem under the 
local constraints as
%
%
\begin{eqnarray}
\hat{Q}_i=b_{i}^{\dag}b_{i}+\sum_{\Gamma}f_{i\Gamma}^{\dag}f_{i\Gamma}=1.
\end{eqnarray}
The expectation value of an operator $\hat{O}$ under the local constraint eq. (2.9)
is given as
%
%
\begin{eqnarray}
\langle \hat{O} \rangle = \lim_{ \{ \lambda_i \} \rightarrow \infty  }
\langle \hat{O} \prod_i\hat{Q}_i \rangle_{\lambda} /\langle  \prod_i\hat{Q}_i \rangle_{\lambda},
\end{eqnarray}
where $\langle \hat{A}  \rangle_{\lambda}$ is calculated in the grand canonical ensemble
for the Hamiltonian $H_{\lambda}$: 
%
%
\begin{eqnarray}
&&\langle \hat{A}  \rangle_{\lambda} \equiv \textrm{Tr}\Bigl[\textrm{e}^{-\beta H_{\lambda}}\hat{A}\Bigr]
/\textrm{Tr}\Bigl[\textrm{e}^{-\beta H_{\lambda}}\Bigr],\\
&& H_{\lambda}=H+\sum_{i}\lambda_i \hat{Q}_i.
\end{eqnarray}
The single-particle Green functions $G_{{\bf k}\sigma}$ for the conduction electron, $B_i$
for the slave boson and $F_{i \Gamma}$ for the pseudo-fermion are essential ingredients. 
Their unperturbed forms are given as follows:
%
%
\begin{align}
G^{0}_{{\bf k}\sigma}(\textrm{i}\omega_n)&=(\textrm{i}\omega_n-\varepsilon_{{\bf k \sigma}})^{-1},\\
B^{0}_{i}(\textrm{i}\nu_n)&=(\textrm{i}\nu_n-\lambda_i)^{-1},\\
F^{0}_{i \Gamma}(\textrm{i}\omega_n)&=(\textrm{i}\omega_n-\lambda_i-E_{i\Gamma})^{-1},
\end{align}
%
%
\begin{figure}[t]
\begin{center}
\includegraphics[scale=0.4]{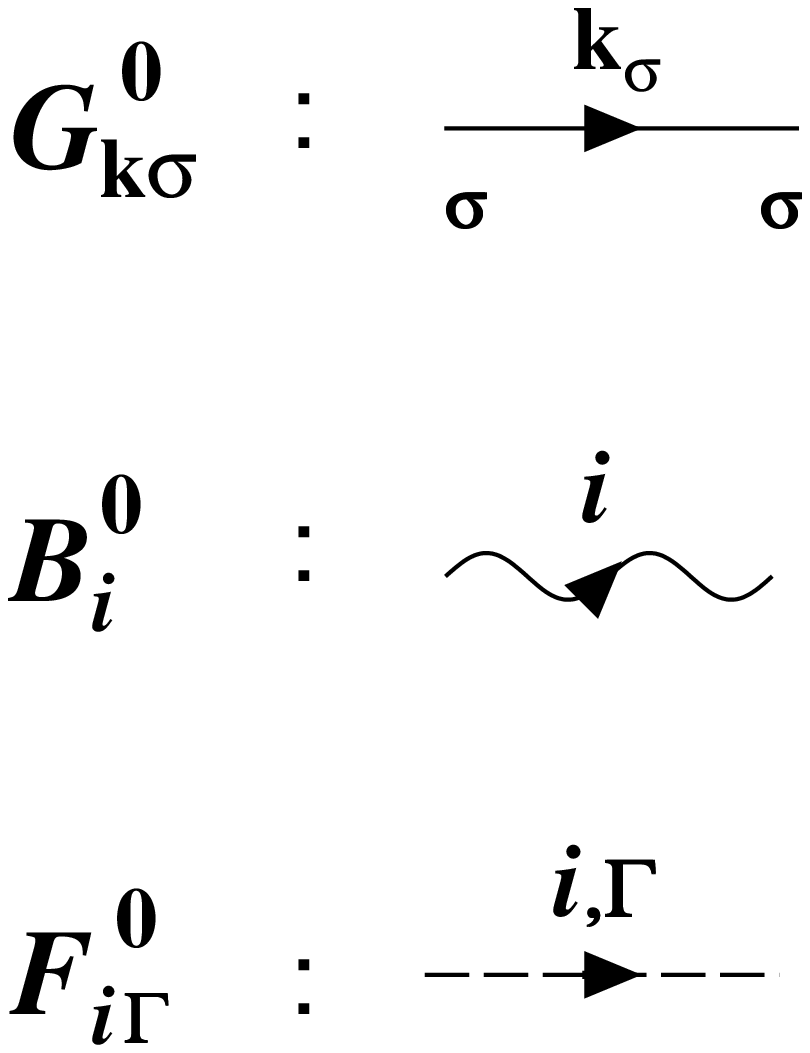}
\end{center}
\caption{Feynman diagrams: A solid line denotes the conduction electron propagator, 
a wavy line the slave-boson propagator, and a dashed line the pseudo-fermion propagator. }
\label{fig2}
\end{figure}
where $\omega_n =(2n+1)\pi T$, and $\nu_n =2n\pi T$ with $n$ being an integer.
Then, the Feynman diagrams are illustrated in Fig. \ref{fig2},
and those for the $cf$-mixing vertices are illustrated in Fig. \ref{fig3}.
The effect of the interaction, eq. (2.4), can be formally incorporated into the Green function
as self-energies as follows:
%
%
\begin{align}
G_{{\bf k}\sigma}(\textrm{i}\omega_n)&=(\textrm{i}\omega_n-\varepsilon_{{\bf k \sigma}}
-\Sigma_{{\bf k}\sigma}(\textrm{i}\omega_n))^{-1},\\
B_{i}(\textrm{i}\nu_n)&=(\textrm{i}\nu_n-\lambda_i-\Pi_i(\textrm{i}\nu_n))^{-1},\\
F_{i \Gamma}(\textrm{i}\omega_n)&=(\textrm{i}\omega_n -E_{i \Gamma} -\lambda_i-
\Sigma^{f}_{i\Gamma}(\textrm{i}\omega_n))^{-1}.
\end{align}
The self-energies $\Sigma_{{\bf k}\sigma}$, $\Pi_i$, and $\Sigma^{f}_{i\Gamma}$ 
are calculated by an extended 1/$N$-expansion \textit{\'{a}} $la$ Nagoya, as discussed below.
\subsection{1/N-expansion \textit{\'{a}} 
$la$ Nagoya for SU(N)-PAM}
We briefly review the past work performed by \=Ono $et\ al.$
as detailed in refs. \citen{Ono1,Ono2}.
The conventional $1/N$-expansion method for 
SU($N$)-PAM, in which the conduction band is prepared for each component of pseudo-fermion 
and the hybridization depends only on {\bf k}, is performed by dividing the k-space and spin-space 
of the conduction electron into $N$ subspace and keeping the total degrees of freedom of the 
conduction electron to be the same as those of the non-interacting conduction band:
%
%
\begin{align}
\{\uparrow,k\}+\{\downarrow,k\}=\{k_1\}+\{k_2\}+\cdots\{ k_m\}+\cdots+\{ k_N\},
\end{align}
%
%
\begin{align}
\sum_{m=-J}^{m=J}\Bigl(\frac{1}{N_L}\sum_{k_m}1\Bigr)=2.
\end{align}
where $N_L$ stands for the number of total lattice sites, $N_L \equiv \sum_{i} 1$.
We assume that the ${k_m}$ state of the conduction electron hybridizes only with the $m$-th state
of the f-electron. Then, the density of states (DOS) of the $m$-th subspace of the conduction electron is
$1/N$ times as small as the total DOS, so that a summation over $k_m$ gives a factor 
of $1/N$. Therefore, we can classify the Feynman diagrams giving the self-energies of the Green functions
in terms of $1/N$. This method may be justified on the basis of the following intuitive argument.

The state of the conduction electron with momentum ${\bf k}$ and spin ${\sigma}$ is expanded around the origin, where
an f-electron is located, into that of the f-electron with the state of angular momentum state $|m \rangle$:
%
%
\begin{align}
c_{\bf{k} \sigma}=\sum_{m}\sum_{k_m} \langle \sigma,{\bf{k}}|m,k_m\rangle c_{k_m}.
\end{align}
Namely, only a restricted part of the degrees of freedom {\bf k} and $\sigma$ of the conduction electron near the Fermi level hybridizes with f-electron in the state $|m\rangle$.
If the f-electron state is specified by $N$ different quantum numbers, the degrees of freedom of conduction
electrons which hybridize with the f-electron with the state $|m\rangle$ are equal to the roughly $1/N$ times
of the total degrees of freedom of the conduction electron,
so that the effective DOS of a relevant conduction electron is reduced by $1/N$ times.
In the lattice system, hybridization between conduction electrons with ${\bf k}$ and $\sigma$ and
the f-electron with $|m\rangle$ at the $i$-th site is given in terms of the matrix element in (2.21) as
%
%
\begin{align}
\langle \sigma,{\bf k}|k_m,m;i\rangle=e^{-i{\bf k}\cdot{\bf R}_i}\langle \sigma,{\bf k}|m,k_m\rangle.
\end{align}
Therefore, the Hamiltonian $H_{cf}$ for hybridization is given as follows:
%
%
%
\begin{figure}[t]
\begin{center}
\includegraphics[scale=0.4]{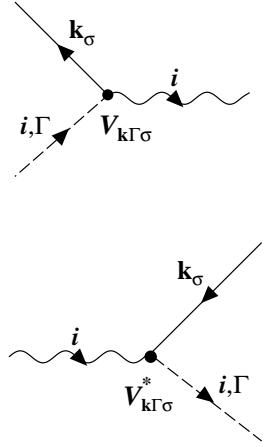}
\end{center}
\caption{Feynman diagrams for cf-hybridization: A solid line denotes the conduction electron propagator, 
a wavy line the slave-boson propagator, and a dashed line the pseudo-fermion propagator. 
These propagators are connected by the cf-hybridization vertex due to $H_{cf}$, eq. (2.4).}
\label{fig3}
\end{figure}
%
%
%
\begin{align}
H_{cf}=&\sum_{i,m}\sum_{\sigma,{\bf k}}\Bigl( \langle \sigma,{\bf k}|H|m;i\rangle c_{{\bf k}
\sigma}^{\dag}f_{im}+\textrm{h.c.}\Bigr),\\
=&\sum_{i,m}\sum_{m',k_{m'}}\sum_{\sigma,{\bf k}}\Bigl(\langle m',k_{m'} | \sigma,{\bf k} \rangle
\langle \sigma,{\bf k}|H|m;i \rangle 
c_{k_{m'}}^{\dag} f_{im}+\textrm{h.c.}\Bigr),\\
=&\sum_{i,m}\sum_{m',k_{m'}}\Bigl( \langle m',k_{m'}|H|m;i \rangle
c_{k_{m'}}^{\dag} f_{im}
+\textrm{h.c.}\Bigr),\\
=&\sum_{i,m}\sum_{m',k_{m'}}\Bigl( 
\frac{1}{\sqrt{N_L}}V\delta_{m,{m'}} \textrm{e}^{-\textrm{i}k_{m'} R_i} c_{k_{m'}}^{\dag}f_{im}
+\textrm{h.c.}\Bigr),\\
=&\frac{1}{\sqrt{N_L}}\sum_{i,m}\sum_{k_m} \Bigl(V  c_{k_m}^{\dag}f_{im}\textrm{e}^{-\textrm{i}k_{m} R_i}+\textrm{h.c.}\Bigr).
\end{align}
Similarly, the Hamiltonian $H_{c}$ for the kinetic term is given as follows:
\begin{align}
H_{c}=&\sum_{\sigma,{\bf k}}\sum_{\sigma',{\bf k}^{'}}\langle \sigma,{\bf k}|H|\sigma',{\bf k'}\rangle c_{{\bf k}
\sigma}^{\dag}c_{{\bf k}^{'}\sigma^{'}},\\
=&\sum_{m,k_m}\sum_{m',k_{m'}}\sum_{\sigma,{\bf k}}\sum_{\sigma',{\bf k}^{'}}
\langle m,k_{m} | \sigma,{\bf k} \rangle
\langle \sigma,{\bf k}|H| \sigma',{\bf k'} \rangle 
\langle \sigma',{\bf k'}|m',k_{m'} \rangle 
c_{k_{m}}^{\dag}c_{k_{m'}},\\
=&\sum_{m,k_{m}}\sum_{m',k_{m'}} \langle m,k_{m}|H|m',k_{m'} \rangle
c_{k_{m}}^{\dag}c_{k_{m'}} ,\\
=&\sum_{m,k_{m}} \varepsilon_{k_m} c_{k_{m}}^{\dag}c_{k_{m}}.
\end{align}
Here, we have replaced the matrix elements $\langle m',k_{m'} |H|m;i\rangle$ in eq. (2.25)
with $V\delta_{m,{m'}} \textrm{e}^{-\textrm{i}k_m R_i}
/\sqrt{N_L}$, and we have replaced $\langle m,k_m |H|m', k_{m'}
\rangle$ in eq. (2.30)
with $\varepsilon_{k_m} \delta_{m,m'}\delta_{k_m,k_{m'}}$.
Then, the rareness of the DOS of the conduction electron for each channel $m$ is inherited in the lattice system. 
In the lattice system, strictly speaking, the conduction electrons described by $c_{k_m}$'s mix with each other 
so that a classification in terms of the index $m$ becomes ambiguous. So the approximation in eqs. (2.26) and (2.31) 
for the conduction electrons should be regarded as an approximation that neglects the hybridization of conduction 
electrons with different indices $m$. Nevertheless, this approximate treatment has been shown to provide reasonable results 
for numerous strongly correlated electron systems even for the case $N=2$\cite{tsuru}.

However, an extension of the conventional SU($N$)-PAM enabling inclusion of CEF splitting $\Delta$ 
causes a few problems: e.g., the pole of $B_i(\nu)$, $E_0$ which corresponds to the Kondo temperature $T_K$,
results in contradiction, i.e., $\lambda+\varepsilon_f-E_0>\lambda+\varepsilon_f $, if $\Delta$ is larger 
than almost twice $T_K$ for $\Delta=0$, i.e., without CEF splitting. In a word, the energy level of the 
coherent state becomes higher than that of the bare f-electron, and the 1/$N$-expansion method fails.
For example, in the case where four degenerate states of the f-electron, i.e., $N=4$, 
are divided into two Kramers doublets by CEF splitting $\Delta$, the correct Kondo 
temperature $T_K^{(2)}$ for $N=2$ is not obtained in the limit $\Delta \rightarrow \infty$.
One of the origins of this difficulty may be that the hybridization between conduction electrons with different 
quantum numbers $m$ is discarded, so that the two components of the conduction electrons are 
independent of each other.
To resolve this problem, we reconsider a generalized PAM in which the conduction electron, 
specified by {\bf k} and $\sigma$, mixes with all the states of f-electrons specified by $m$.
\subsection{Extended 1/N-expansion method for generalized PAM}
We now apply the 1/$N$-expansion method to the generalized PAM introduced
in \S 2.1.
Namely, we extend a conventional 1/$N$-expansion method in such a way that
the condition of eq. (2.20) is replaced by the following conditions.
%
%
\begin{align}
&\frac{1}{N_L}\sum_{k_\sigma}1=1=O(1/N),\\
&\sum_{\sigma}1=2=O(N),\\
&\frac{1}{N_L}\sum_{\sigma}\sum_{k_\sigma}1=2=O(1).
\end{align}
Then, the degeneracy $N$ is given by just that of the real spin of the conduction electrons: i.e., $N=2$.
We use this condition hereafter in the present paper.
As shown below, the rule of classification of the Feynman diagrams with respect to $1/N$ in the SU($N$)-PAM model can be 
generalized in the present model that has only spin degeneracy $N=2$. The validity of using the $1/N$-expansion method \textit{\'{a}} 
$la$ Nagoya in the case $N=2$ is supported by the results of previous studies that treated a single-band PAM 
with $N=2$ and gave qualitatively correct physical behavior in both the high- and low-energy regions\cite{tsuru}.
In particular, it has been shown that higher order corrections in $1/N$ do not change the results qualitatively\cite{tsuru2}.
On the basis of this observation, we use this $1/N$-expansion as a method of taking into account the correlation effect of f-electrons 
in the present model.

\begin{figure}[t]
\begin{center}
\includegraphics[scale=0.35]{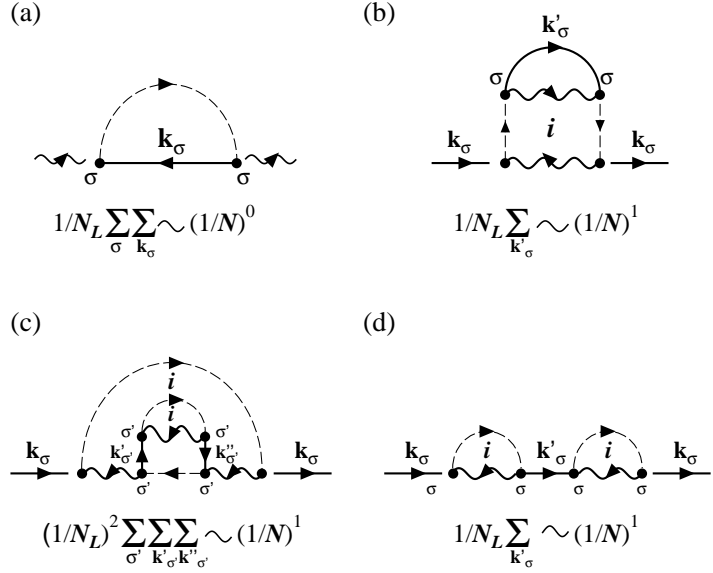}
\end{center}
\caption{Example of diagrams showing contributions to the self-energy:
the lowest order of the self-energy for the slave boson (a)
and the next order examples of the self-energy for the 
conduction electron (b,c,d) in power of ($1/N$).}
\label{fig4}
\end{figure}

\begin{figure}[t]
\begin{center}
\includegraphics[scale=0.35]{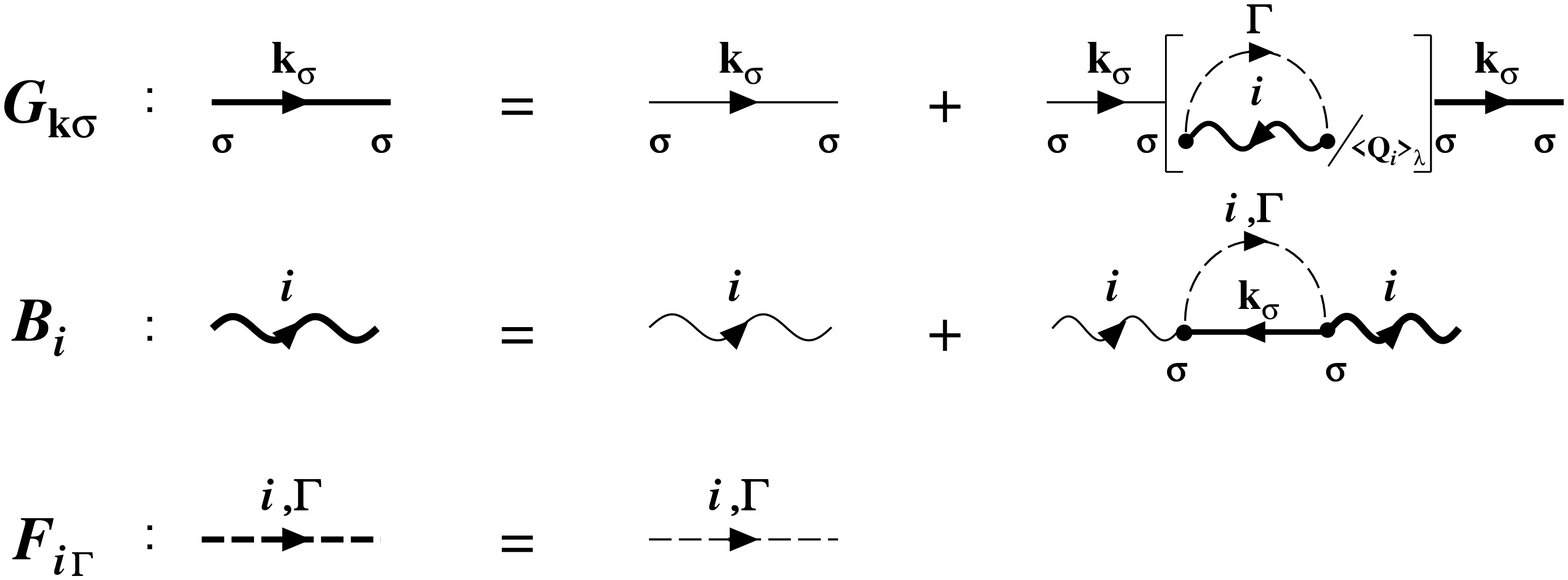}
\end{center}
\caption{Diagrammatic representation of Dyson equations for 
single particle Green function within accuracy of $(1/N)^0$. 
Thin solid, wavy, and dashed lines represent the Green function 
of the conduction electron, slave-boson, and pseudo-fermion, respectively. 
Thick lines denote the renormalized Green function for each particle.}
\label{fig5}
\end{figure}

The self-energies of the conduction electron $\Sigma_{{\bf k}\sigma}$, in eq. (2.16), the slave-boson $\Pi_i$,
in eq. (2.17), and the pseudo-fermion $\Sigma^{f}_{i\Gamma}$, in eq. (2.18), are determined 
for each order of $1/N$.
To classify the diagrams by each order of $(1/N)$, we use the condition in eqs. (2.32) $\sim$ (2.34). 
We illustrate several examples of the self energy in Fig. \ref{fig4}.
In our extended formulation of the $1/N$-expansion, the diagram of Fig. \ref{fig4}(a) is of $O(1/N)^0$, 
because the summation with respect to both ${{\bf k}_\sigma}$ and $\sigma$ is performed as seen in eq. (2.34). 
On the other hand, those of Fig. \ref{fig4}(b), Fig. \ref{fig4}(c) and Fig. \ref{fig4}(d) are of $O(1/N)^1$
because one summation over ${\bf k}_\sigma$ remains. 
For example, in Fig. \ref{fig4}(b), the summation
$(1/N_L)^2 \sum_{i}\sum_{k'_\sigma}$ is performed, so that this diagram gives the order of
$(1/N_L) \sum_{k'_\sigma}\sim O(1/N)$.
Similarly, in Fig. \ref{fig4}(c), the summation $(1/N_L)^3\sum_{i}\sum_{\sigma'}\sum_{k'_{\sigma'}}\sum_{k''_{\sigma'}}$
is performed. This also gives $(1/N_L)^2\sum_{\sigma'}\sum_{k'_{\sigma'}}\sum_{k''_{\sigma'}} \sim (1/N_L)\sum_{k''_{\sigma'}}\sim O(1/N)$.
%
%
Then, for the leading order in $1/N$, the Dyson equation for the conduction electron, 
the slave-boson, and pseudo-fermion Green function are given by the Feynman diagrams illustrated 
in Fig. \ref{fig5}. It is noted that $\Sigma^{f}_{i\Gamma}$ vanishes at this order of approximation.
Hereafter, we discuss the problem within the accuracy to the leading order of $1/N$, although there 
is no difficulty in taking into account higher order corrections in $1/N$.
\subsection{Dyson equation to the leading order in $(1/N)^0$}
\begin{figure}[t]
\begin{center}
\includegraphics[scale=0.35]{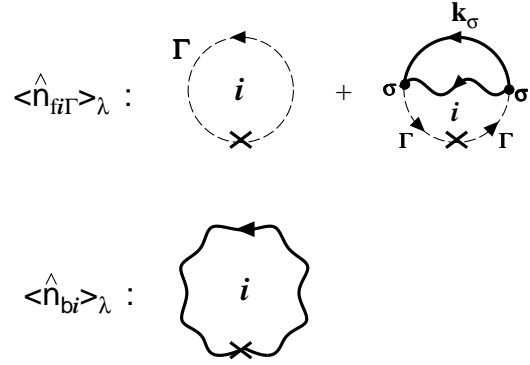}
\end{center}
\caption{Feynman diagrams for $\langle \hat{n}_{fi\Gamma} \rangle_{\lambda}$, 
and $\langle \hat{n}_{bi} \rangle_{\lambda}$ within the order of $(1/N)^0$. }
\label{fig6}
\end{figure}

When treating the problem under zero magnetic field or
without magnetic order, we can neglect the off-diagonal self-energy part, 
with respect to the spin, of the conduction electrons, because of the properties
of the Clebsch-Gordan coefficient. Indeed, from the relation of spherical harmonic
function $\{ Y_{l}^{m}(\Omega_{{k}})   \}^{*}=(-1)^{m} Y_{l}^{-m}(\Omega_{{ k}})$, 
we can find
%
%
%
%
%
%
\begin{align}
&\sum_{\Gamma\pm}V_{{\bf k}\Gamma \sigma}^{*} V_{{\bf k}\Gamma \bar{\sigma}}=0,\\
&\sum_{\Gamma\pm}|V_{{\bf k}\Gamma \sigma}|^2=
\sum_{\Gamma\pm}|V_{{\bf k}\Gamma \bar{\sigma}}|^2,
\end{align}
where $\bar{\sigma}=-\sigma$ and ${\sum_{\Gamma\pm}}$ means the summation with respect to the Kramers
doublets characterized with $| \Gamma \rangle$ = $| \pm 1\rangle$, $| \pm 2\rangle$, and $| \pm 3\rangle$. 
For $\Gamma= | \pm 1\rangle$, eq. (2.35) means that
$V_{{\bf k}| + 1\rangle \sigma}^{*} V_{{\bf k}| + 1\rangle \bar{\sigma}}
+V_{{\bf k}| - 1\rangle \sigma}^{*} V_{{\bf k}| - 1\rangle \bar{\sigma}}
=0$ is satisfied.
The analytic form of the relevant self-energy parts are given by
%
%
\begin{align}
\Sigma_{{\bf k}\sigma}(\textrm{i}\omega_n)&=
\lim_{\lambda_i \rightarrow \infty}\Bigl[-\sum_{\Gamma\pm}|V_{{\bf k}\Gamma \sigma}|^{2} T\sum_{\nu_n}
B_i(\textrm{i}\nu_n) F^{0}_{i \Gamma}(\textrm{i}\omega_n+\textrm{i}\nu_n)/\langle \hat{Q}_i \rangle_{\lambda}\Bigr],\\
\Pi_{i}(\textrm{i}\nu_n)&=\frac{1}{N_L}\sum_{\Gamma\pm} \sum_{\sigma}\sum_{{\bf k}_\sigma}|V_{{\bf k}\Gamma \sigma}|^{2}
T\sum_{\omega_n} G_{{\bf k}\sigma}(\textrm{i}\omega_n) F^{0}_{i\Gamma}(\textrm{i}\omega_n+\textrm{i}\nu_n),\\
\langle \hat{Q}_{i}\rangle_{\lambda}&=\sum_{\Gamma\pm}\langle \hat{n}_{fi\Gamma}\rangle_{\lambda}+\langle \hat{n}_{bi}\rangle_{\lambda}.
\end{align}
Here, $\langle \hat{n}_{fi\Gamma}\rangle_{\lambda} \equiv \langle f_{i\Gamma}^{\dag} f_{i\Gamma}\rangle_{\lambda}$
and $\langle \hat{n}_{bi}\rangle_{\lambda} \equiv \langle b_{i}^{\dag} b_{i}\rangle_{\lambda}$ are given
by the diagrams shown in  Fig. \ref{fig6} whose analytic forms are given as follows:
%
%
\begin{align}
\langle \hat{n}_{fi\Gamma}\rangle_{\lambda}
=&T\sum_{\omega_n}F^{0}_{i\Gamma}(\textrm{i}\omega_n) \\
&-\frac{1}{N_L}\sum_{\sigma}\sum_{{\bf k}_\sigma}|V_{{\bf k}\Gamma \sigma}|^2 T^2 
\sum_{\omega_n}\sum_{\nu_n}
F^{0}_{i\Gamma}(\textrm{i}\omega_n+\textrm{i}\nu_n)^2 G_{{\bf k}\sigma}(\textrm{i}\omega_n)B_i(\textrm{i}\nu_n),\\
\langle \hat{n}_{bi}\rangle_{\lambda}=&-T\sum_{\nu_n}B_i(\textrm{i}\nu_n).
\end{align}
Taking the summation with respect to the Matsubara frequency in eqs. (2.37) and (2.38), 
and performing the analytic continuation $\textrm{i}\omega_n \rightarrow \omega_{+}\equiv \omega+\textrm{i}0_+$,
we obtain the imaginary part and real part of the self-energies as follows:
%
%
\begin{align}
\textrm{Im}\Sigma_{{\bf k}\sigma}(\omega_{+})=&\lim_{\lambda_i \rightarrow \infty}\sum_{\Gamma\pm} |V_{{\bf k}\Gamma \sigma}|^2
\Bigl( 1+\textrm{e}^{\beta \omega} \Bigr)\textrm{e}^{-\beta(\lambda_i+E_{i \Gamma})}\nonumber\\
&\times\textrm{Im}B_i(-\omega+\lambda_i+E_{i\Gamma}+\textrm{i}0_+)/\langle \hat{Q}_i \rangle_{\lambda},\\
\textrm{Re}\Sigma_{{\bf k}\sigma}(\omega_{+})=&-\frac{1}{\pi}P\int_{-\infty}^{\infty} \textrm{d} \omega' \frac{1}{\omega-\omega' +\textrm{i} 0_+}
\textrm{Im}\Sigma_{{\bf k}\sigma}({\omega'}_{+}),\\
\textrm{Im}\Pi_{i}(\omega_{+})=&\frac{1}{N_L}\sum_{\Gamma\pm}\sum_{\sigma}\sum_{{\bf k}_\sigma} |V_{{\bf k}\Gamma \sigma}|^2
f(-\omega+\lambda_i+E_{i \Gamma})\nonumber\\
&\times \textrm{Im}G_{{\bf k}\sigma}(-\omega+\lambda_i+E_{i \Gamma}+\textrm{i}0_+),\\
\textrm{Re}\Pi_{i}(\omega_{+})=&-\frac{1}{\pi}P\int_{-\infty}^{\infty} \textrm{d}\omega' \frac{1}{\omega-\omega' +\textrm{i} 0_+}
\textrm{Im}\Pi_{i}({\omega'}_{+}),
\end{align}
where $f(\omega) \equiv (e^{\beta \omega}+1)^{-1}$ is the Fermi distribution function
($\beta \equiv 1/T$).
The real part of the self-energies are given by the Kramers-Kr\"{o}nig relation from the corresponding imaginary part 
of the self-energy. Of course, we can calculate $\Sigma_{{\bf k}\sigma}$ and $\Pi_{i}$ analytically without separating
the real and imaginary parts. However, we express them in such a way for convenience of the numerical calculation.

Here, we introduce characteristic parameters $E_0$ and $a$. From eqs. (2.17), (2.45), and (2.46), 
we can find the following  relation at low temperature ($T\lesssim E_0/10$, as shown later).
%
%
\begin{align}
-\frac{1}{\pi}\textrm{Im}{B}_i(\omega_{+})=&a(T)\cdot\delta(\omega+\lambda_i+ \varepsilon_f+ E_0(T))
+C(\omega+\lambda_i+\varepsilon_f),
\end{align}
where $C(\omega+\lambda_i+\varepsilon_f)$ is a broad continuous function that is non-vanishing only in the region of
$\omega > 0$, and the binding energy $E_0(T)$ and the residue $a(T)$ of the slave-boson are determined through
the coupled relations
%
%
\begin{align}
E_0(T)&=\varepsilon_f-\textrm{Re}\Pi_i(\lambda_i + \varepsilon_f-E_0(T)),\\
\frac{1}{a}&=1-\frac{\textrm{d}}{\textrm{d}\omega} (\textrm{Re}\Pi_i(\omega))|_{\omega=\lambda_i+\varepsilon_f-E_0(T)}.
\end{align}
%
%
The parameter $E_0 \equiv E_0(0)$ corresponds to the Kondo temperature $T_K$ in the case of 
impurity Anderson model\cite{kuro}. To express a series of equations more concisely in the 
actual numerical calculation, we introduce the following notations:
$\langle \hat{Q}_i \rangle_{\lambda}=\textrm{e}^{-\beta(\lambda_i+\varepsilon_f-E_0(T))}\langle \hat{Q} \rangle$, 
$\bar{B}(\omega)=B_i(\omega+\lambda_i+\varepsilon_f)$ and 
$\bar{\Pi}(\omega)=\Pi_i(\omega+\lambda_i+\varepsilon_f)$.
Then, we can rewrite eqs. (2.37)$\sim$(2.42) as follows:
%
%
\begin{align}
\textrm{Im}\Sigma_{{\bf k}\sigma}(\omega_{+})=&\sum_{\Gamma\pm} |V_{{\bf k}\Gamma \sigma}|^2
\Bigl( 1+\textrm{e}^{\beta \omega} \Bigr) \textrm{e}^{-\beta[E_0(T)+E_{i\Gamma}-\varepsilon_f]}\nonumber\\
&\times \textrm{Im}\bar{B}
(-\omega+E_{i\Gamma}-\varepsilon_f+\textrm{i}0_+)/\langle \hat{Q} \rangle,\\
\textrm{Im}\bar{\Pi}(\omega_{+})=&\frac{1}{N_L}\sum_{\Gamma\pm}\sum_{\sigma}\sum_{{\bf k}_\sigma}
|V_{{\bf k}\Gamma \sigma}|^2 f(-\omega+E_{i \Gamma}-\varepsilon_f)\nonumber\\
&\times \textrm{Im}G_{{\bf k}\sigma}
(-\omega+E_{i \Gamma}-\varepsilon_f+\textrm{i}0_+),
\end{align}
and
\begin{align}
\langle \hat{n}_{fi\Gamma} \rangle_{\lambda}
=&\textrm{e}^{-\beta(\lambda_i+\varepsilon_f-E_0(T))}\times \Bigl[ \textrm{e}^{-\beta[E_0(T)+E_{i \Gamma}-\varepsilon_f]}\nonumber\\
&+\sum_{\sigma}\sum_{{\bf k}_\sigma}\frac{|V_{{\bf k}\Gamma \sigma}|^2}{N_L}
\int \textrm{d} \omega \ f(\omega)
\frac{-1}{\pi}\textrm{Im}G_{{\bf k}\sigma}(\omega_+)\nonumber \\
&\times \int \textrm{d} \omega'\  \textrm{e}^{-\beta(E_0(T)+\omega')}
\frac{-1}{\pi}\textrm{Im}\bar{B}(\omega'_+)
\cdot \frac{1}{(\omega+\omega'-E_{i\Gamma}+\varepsilon_f)^2}\nonumber\\
&+\sum_{\sigma}\sum_{{\bf k}_\sigma}\frac{|V_{{\bf k}\Gamma \sigma}|^2}{N_L}
\int \textrm{d} \omega \ \textrm{e}^{-\beta (E_0(T)+E_{i\Gamma}-\varepsilon_f)}  \frac{-1}{\pi}\textrm{Im}G_{{\bf k}\sigma}(-\omega+\textrm{i}0_+) \nonumber\\
&\times f(\omega)\cdot \bigl\{ \frac{-1}{T}\bar{B}(\omega+E_{i\Gamma}-\varepsilon_f)
+\frac{\textrm{d} \bar{B}(\omega')}{\textrm{d}\omega'}\Bigr|_{\omega'=\omega+E_{i\Gamma}-\varepsilon_f} \bigr\} \Bigr],\\
\langle \hat{n}_{bi} \rangle_{\lambda} =&\textrm{e}^{-\beta(\lambda_i+\varepsilon_f-E_0(T))}
\times  \int \textrm{d} \omega \ \textrm{e}^{-\beta(\omega +E_0(T))}\frac{-1}{\pi}\textrm{Im}\bar{B}(\omega_+).
\end{align}
The validity of the approximate eq. (2.47) in the low-temperature region, or eq. (2.54) below, 
for $\bar{B}$ is shown in the next subsection, while in high-temperature region ($T>E_0$)
eq. (2.47) ceases  to be valid. 
%
%
\begin{align}
-\frac{1}{\pi}\textrm{Im}{\bar{B}}(\omega_{+})=&a(T)\cdot\delta(\omega+ E_0(T))+C(\omega).
\end{align}
To extend this formulation to a finite temperature region, we must use the Dyson equation self-consistently (See  $\S 4$).

Finally we calculate the average number $n_c$ of conduction electrons and $n_f$ of f-electrons per site:
%
%
\begin{align}
n_c &\equiv \frac{1}{N_L}\sum_{\sigma}\sum_{{\bf k}_\sigma}\Bigl[T\sum_{\omega_n}
G_{{\bf k}\sigma}(\textrm{i}\omega_n)e^{\textrm{i}\omega_n 0+}\Bigr],\\
n_f &\equiv \lim_{\lambda_i \rightarrow \infty}\Bigl[ \sum_{\Gamma\pm} \langle \hat{n}_{f i \Gamma}
\rangle_{\lambda}/\langle \hat{Q}_i \rangle_{\lambda}\Bigr].
\end{align}
With the use of eqs. (2.37) and (2.41), we obtain
%
%
\begin{align}
n_f=&\lim_{\lambda_i \rightarrow \infty}
\sum_{\Gamma\pm}\Bigl[T\sum_{\omega_n}F_{i\Gamma}^0 (\textrm{i}\omega_n)/\langle \hat{Q}_i \rangle_{\lambda}\Bigr]\\
&+\frac{1}{N_L}\sum_{\sigma}\sum_{{\bf k}_\sigma}\Bigl[ -T\sum_{\omega_n}
G_{{\bf k}\sigma}(\textrm{i}\omega_n)
\frac{\textrm{d}}{\textrm{d}(\textrm{i}\omega_n)}\Sigma_{{\bf k}\sigma}(\textrm{i}\omega_n)\Bigr].
\end{align}
Hence, the total number of electrons per site is given by
%
%
\begin{align}
n_c+n_f=&\lim_{\lambda_i \rightarrow \infty}
\sum_{\Gamma\pm}\Bigl[T\sum_{\omega_n}F_{i\Gamma}^0 (\textrm{i}\omega_n)/\langle \hat{Q}_i \rangle_{\lambda}\Bigr]\nonumber\\
&+\frac{1}{N_L}\sum_{\sigma}\sum_{{\bf k}_\sigma}\Bigl[T\sum_{\omega_n}
\frac{\textrm{d}}{\textrm{d}(\textrm{i}\omega_n)}
\log G^{-1}_{{\bf k}\sigma}(\textrm{i}\omega_n)\Bigr],\\
=&\sum_{\Gamma\pm}\textrm{e}^{-\beta(E_0(T)+E_{i\Gamma}-\varepsilon_f)}/\langle \hat{Q}\rangle\nonumber\\
&+
\frac{1}{N_L}\sum_{\sigma}\sum_{{\bf k}_\sigma}\Bigl[T\sum_{\omega_n} 
\frac{\textrm{d}}{\textrm{d}(\textrm{i}\omega_n)}
\log G^{-1}_{{\bf k}\sigma}(\textrm{i}\omega_n)\Bigr].
\end{align}
At $T=0$, this relation is simplified as
%
%
\begin{align}
n_c+n_f=&\frac{1}{N_L}\sum_{\sigma}\sum_{{\bf k}_\sigma}\Bigl[T\sum_{\omega_n}
 \frac{\textrm{d}}{\textrm{d}(\textrm{i}\omega_n)}
\log G^{-1}_{{\bf k}\sigma}(\textrm{i}\omega_n)\Bigr].
\end{align}
Furthermore, we find $\textrm{Im}\Sigma_{{\bf k}\sigma}(0_+)=0$ at $T=0$ from eqs. (2.50) and (2.54), 
so that we obtain the total number $n \equiv n_c+n_f$ as
%
%
\begin{align}
n=n_c+n_f=\frac{1}{N_L}\sum_{\sigma}\sum_{{\bf k}_\sigma}\theta(-\varepsilon_{{\bf k}\sigma}-\Sigma_{{\bf k}\sigma}(0)).
\end{align}
The total number is determined by the volume of the $k$-space enclosed by the Fermi surface.
In a word, the Landau-Luttinger sum-rule holds at $T=0$\cite{Lut} . 
%
%
%
\section{Physical properties of Generalized Periodic Anderson Model at $T=0$}
In this section, we study the properties of the generalized PAM (2.1)$\sim$(2.4) 
in the limit of $T \rightarrow 0$, where eq. (2.54) (or eq. (2.47)) is a valid 
approximation. (See  $\S\ 3.2$)
With the use of eq. (2.54), we manipulate eqs. (2.52) and (2.53) as follows. 
From eqs. (2.52) $\sim$ (2.54)
%
%
\begin{align}
\langle \hat{n}_{fi\Gamma}\rangle_{\lambda}=&\textrm{e}^{-\beta(\lambda_i+\varepsilon_f-E_0)}\nonumber\\
&\times \Bigl[\textrm{e}^{-\beta(E_0+E_{i\Gamma}-\varepsilon_f)}\nonumber\\
&+\sum_{\sigma}\sum_{\bf{k}_\sigma}\frac{a|V_{{\bf k}\Gamma\sigma}|^2}{N_L}\int \textrm{d}\omega
\ \frac{-1}{\pi}\textrm{Im}G_{\bf{k}\sigma}(\omega_+)\cdot
\frac{f(\omega)}{(\omega-E_0-E_{i\Gamma}+\varepsilon_f)^2}\nonumber\\
&+\sum_{\sigma}\sum_{\bf{k}_\sigma}\frac{|V_{{\bf k}\Gamma\sigma}|^2}{N_L}
\textrm{e}^{-\beta E_0}\int \textrm{d}\omega'\ \textrm{e}^{-\beta \omega'} C(\omega')\nonumber\\
&\times\int \textrm{d}\omega\ 
\frac{-1}{\pi}\textrm{Im}G_{\bf{k}\sigma}(\omega_+)\cdot\frac{f(\omega)}{(\omega+\omega'-E_{i\Gamma}+\varepsilon_f)^2}
\Bigr],\\
\langle \hat{n}_{bi}\rangle_{\lambda}=&\textrm{e}^{-\beta(\lambda_i+\varepsilon_f-E_0)}\times
\Bigl[
a+\textrm{e}^{-\beta E_0}\int\textrm{d}\omega\ \textrm{e}^{-\beta \omega}C(\omega)
\Bigr],
\end{align}
where $a\equiv a(0)$. At $T=0$, exp($-\beta E_0$) vanishes for $E_0 >0$, so that
%
%
\begin{align}
\langle \hat{n}_{fi\Gamma}\rangle_{\lambda}=&\textrm{e}^{-\beta(\lambda_i+\varepsilon_f-E_0)}
\sum_{\sigma}\sum_{\bf{k}_\sigma}\frac{a|V_{{\bf k}\Gamma\sigma}|^2}{N_L}\int 
\textrm{d}\omega
\ \frac{-1}{\pi}\textrm{Im}G_{\bf{k}\sigma}(\omega_+)\nonumber\\
&\times \frac{f(\omega)}{(\omega-E_0-E_{i\Gamma}+\varepsilon_f)^2},\\
\langle \hat{n}_{bi}\rangle_{\lambda}=&a\textrm{e}^{-\beta(\lambda_i+\varepsilon_f-E_0)}.
\end{align}
Similarly, when we use eqs. (2.37), (2.43), and (2.44) at $T=0$,
the continuum part of the spectrum of the slave-boson Green function $C(\omega)$ in eq. (2.54) can also 
be neglected, because the continuum part is smaller than the contribution from a pole by a 
factor of exp($-\beta E_0$).
As a result the self-energy of the conduction electron, eq. (2.37), and $\langle \hat{Q}_i \rangle_{\lambda}$,
(2.39), are given as
%
%
\begin{align}
\langle \hat{Q}_i \rangle_{\lambda}
&=a (1/a-1) \textrm{e}^{-\beta(\lambda_i + \varepsilon_f -E_0)} +a
\textrm{e}^{-\beta(\lambda_i + \varepsilon_f -E_0)},\nonumber\\
&=\textrm{e}^{-\beta(\lambda_i + \varepsilon_f -E_0)},\\
\Sigma_{{\bf k}\sigma}(\omega_{+})&=\sum_{\Gamma\pm} \frac{a  |V_{{\bf k}\Gamma \sigma}|^2}{\omega_{+} -E_0-E_{i\Gamma}+\varepsilon_f},
\end{align}
where we have used the relations  eqs. (2.38) and (2.49).
Here, we note that the self-energy $\Sigma_{{\bf k}\sigma}(\omega)$ of the 
conduction electrons does not have an imaginary part at $\omega=\mu=0$.
Hence a quasi-particle band is formed and the system behaves as a Fermi liquid, so that the Landau-Luttinger sum-rule holds, 
as seen in the previous section.
From eq. (2.7), we rewrite eq. (3.6) as follows:
%
%
\begin{align}
\textrm{Re}\Sigma_{{\bf k}\sigma}(\omega_{+} )=&\sum_{\pm}
\Bigl[\frac{a |V_{{\bf k}|\pm3 \rangle \sigma}|^2}{\omega_{+} -E_0}
+ \frac{a |V_{{\bf k}|\pm2\rangle \sigma}|^2}{\omega_{+} -E_0-\Delta_1}
+ \frac{a |V_{{\bf k}|\pm1\rangle \sigma}|^2}{\omega_{+} -E_0-\Delta_2}\Bigr],
\end{align}
where, $\sum_{\pm}$ means the summation over the Kramers doublets.
We neglect the k-dependence of $V_{{\bf k}\Gamma\sigma}$, and we assume the summation for
$V_{{\bf k}\Gamma \sigma}$ over each of the Kramers doublets specified by $\Gamma$ to be constant, i.e.,
${\sum_{\pm}}|V_{{\bf k}\Gamma \sigma}|^2
={\sum_{\pm}}|V_{\Gamma \sigma}|^2=V^2$ for simplicity. 
However, this assumption does not qualitatively change the physical properties.
Then, we obtain
%
%
\begin{align}
\textrm{Re}\Sigma_{{\bf k}\sigma}(\omega_{+})=&
\Bigl[ \frac{aV^2}{\omega_{+} -E_0}
+\frac{aV^2}{\omega_{+} -E_0-\Delta_1}+ \frac{aV^2}{\omega_{+} -E_0-\Delta_2} \Bigr],\\
G_{{\bf k}\sigma}(\omega_{+})=&\frac{1}{\omega_+ -\varepsilon_{{\bf k} \sigma}-\Sigma_{{\bf k}\sigma}(\omega)},\nonumber\\
=&\sum_{j=1}^{4} \frac{A_{\bf k}^j}{\omega_{+}-\alpha_{{\bf k}\sigma}^j},
\end{align}
where $\alpha_{{\bf k}\sigma}^j$ and $A_{\bf k}^j$, defined by the pole and residue
of the Green function of the conduction electron, must satisfy the following equations
%
%
\begin{align}
&{\alpha_{{\bf k}\sigma}^{j}}-\varepsilon_{{\bf k}\sigma}-\textrm{Re}\Sigma_{{\bf k}\sigma}
({\alpha_{{\bf k}\sigma}^{j}})=0,
\end{align}
and
%
%
\begin{eqnarray}
A_{\bf k}^{j}=\frac{(\alpha_{{\bf k}\sigma}^j-E_0)(\alpha_{{\bf k}\sigma}^j-E_0-\Delta_1)(\alpha_{{\bf k}\sigma}^j-E_0-\Delta_2)}
{(\alpha_{{\bf k}\sigma}^j-\alpha_{{\bf k}\sigma}^{j+1})(\alpha_{{\bf k}\sigma}^j-\alpha_{{\bf k}\sigma}^{j+2})
(\alpha_{{\bf k}\sigma}^j-\alpha_{{\bf k}\sigma}^{j+3})},
\end{eqnarray}
where ${\alpha_{{\bf k}\sigma}^{j}}={\alpha_{{\bf k}\sigma}^{j+4}}$.
The physical meaning of $\alpha_{{\bf k}\sigma}^j$ is the quasi-particle dispersion of 
the Fermi liquid, because the conduction electrons hybridize with
the physical f-electron  through the self-energy $\Sigma_{{\bf k}\sigma}$. 
$E_0$ and $a$ in eq. (3.8) are determined by eqs. (2.48) and (2.49).
The total number $n$ $(= n_c+n_f)$ per site is determined by eqs. (2.55) and (2.56).
%
%
\begin{align}
&E_0-\varepsilon_f= \frac{1}{N_L}\sum_{j=1}^{4}\sum_{\Gamma\pm}\sum_{\sigma}\sum_{{\bf k}_\sigma}
\frac{A_{\bf k}^j |V_{\Gamma \sigma}|^2 f(\alpha_{{\bf k}\sigma}^{j})}{E_0+E_{i\Gamma}-\varepsilon_f-\alpha_{{\bf k}\sigma}^{j}},\\
&\frac{1}{a}=1+\frac{1}{N_L} \sum_{j=1}^{4}\sum_{\Gamma\pm}\sum_{\sigma}\sum_{{\bf k}_\sigma}
\frac{A_{\bf k}^j |V_{\Gamma \sigma}|^2 f(\alpha_{{\bf k}\sigma}^{j})}{(E_0 +E_{i\Gamma}-\varepsilon_f - \alpha_{{\bf k}\sigma}^{j})^2},\\
&n=n_c+n_f=\frac{1}{N_L}\sum_{j=1}^{4}\sum_{\sigma}\sum_{{\bf k}_\sigma} f(\alpha_{{\bf k}\sigma}^{j})A_{\bf k}^{j} + (1-a).
\end{align}
Here, $n_f$ is also derived from $1-a$ because of local constraints eq. (2.9), 
where $a$ is defined as a residue of the slave-boson Green function in eq. (2.54).
At $T=0$, the residue $a$ is equivalent to the number of $f^0$-states per site: i.e., $n_f=1-a$.
By solving a set of self-consistent equations (3.12)$\sim$(3.14), 
we can obtain physical quantities at $T=0$. The solution derived from the above equations is consistent
with that of slave-boson mean field theory at $T=0$.
In the following subsections, we show the numerical results.
%
%
\begin{figure}[t]
\begin{center}
\includegraphics[scale=0.45]{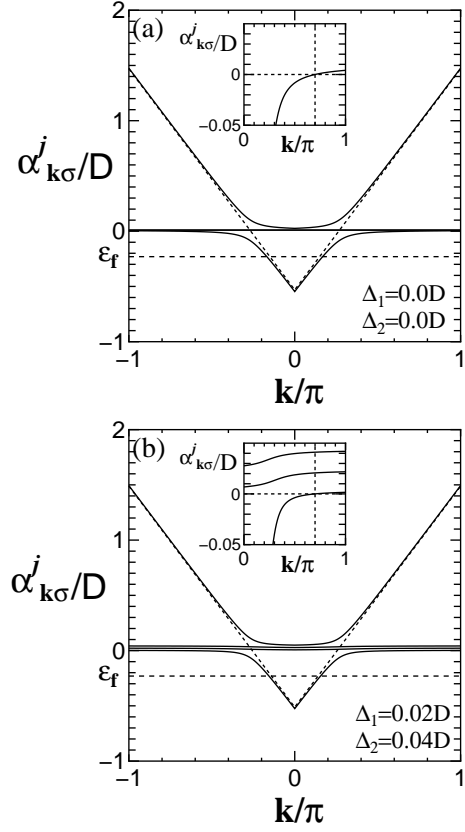}
\end{center}
\caption{Schematic structure of the renormalized bands $\alpha_{{\bf k}\sigma}^{j}$ for the
Fermi level set to zero. $\varepsilon_f$ is the atomic f-level measured from the 
Fermi level.
$\Delta_1$ and $\Delta_2$ are CEF splittings ($ \Delta_1 \le \Delta_2 $). }
\label{fig7}
\end{figure}
%
%
%
%
%
\subsection{Dispersion of renormalized band}
We assume that the dispersion of the bare conduction electrons is linear in any direction,
so that the DOS of the conduction electron 
per spin is approximated as follows:
%
%
\begin{align}
\rho_{\sigma}(\omega)\equiv -\frac{1}{\pi}\frac{1}{N_L}\sum_{{\bf k}_\sigma}\textrm{Im}G^0_{{\bf k}\sigma}(\omega_{+}),
\end{align}
\begin{align}
\rho_{\sigma}(\omega)= \left\{
                   \begin{array}{@{\,}ll}
                   \rho_0=\frac{1}{2D} & \mbox{($-D \le \omega \le  D$)},\\
                   0      & \mbox{(otherwise)}.
                   \end{array}
                 \right.
\end{align}
%
%
\begin{figure}[t]
\begin{center}
\includegraphics[scale=0.45]{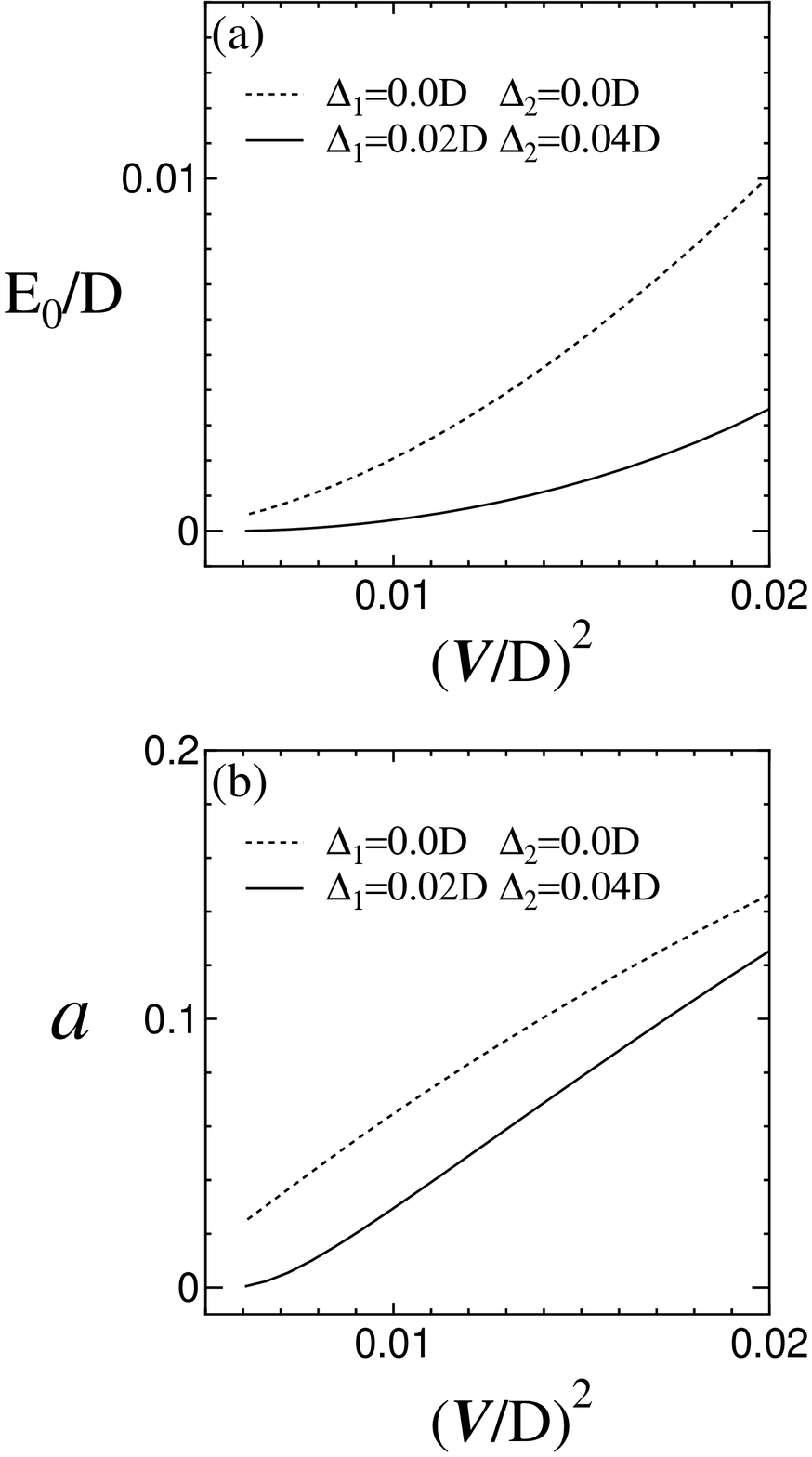}
\end{center}
\caption{(a) $E_0$ vs $V^2$, and (b) $a$ vs $V^2$
for $\Delta_1=\Delta_2=0.0D$ (dotted line), 
and $\Delta_1=0.02D$ $\Delta_2=0.04D$ (solid line).}
\label{fig8}
\end{figure}
We adopt the following parameters; $D=1$, $n \equiv n_c+n_f=1.4$, $V^2=0.02D^2$, $\varepsilon_f^{(0)}=-0.7D$, 
where $\varepsilon_f^{(0)}$ is defined as the energy level of f-electrons measured from the center of the 
conduction electron band, and $\varepsilon_f \equiv \varepsilon_f^{(0)} -\mu$.
Although one might think that the parameter $V$ has too small a value, the binding energy of the slave-boson $E_0$ 
is found to be $E_0$ $\sim0.01 D$ in the case of $\Delta_1=\Delta_2=0$ under these parameters.

We show the dispersion of renormalized bands $\alpha_{{\bf k}}^j$, given by the solution of (3.10),
for two cases of $\Delta_1$ and $\Delta_2$ in Fig. \ref{fig7}(a) and (b), where
the Fermi level is set to be zero. 
In the absence of CEF splitting (Fig. \ref{fig7}(a)), the conduction electron mixes
only with two linear combinations of 6-fold degenerate
states of f-electrons, so that such a band consisting of two linear combinations
is renormalized, and others remain un-renormalized.
On the other hand, in the case of Fig. \ref{fig7}(b) where CEF splitting exists,
the conduction band is renormalized by hybridizing with each f-level, so that the 
dispersion of the renormalized band is divided into four bands.
The insets of Fig. \ref{fig7}(a) and (b) show that the Landau-Luttinger sum-rule holds
because the Fermi wavenumber is fixed as $k_F=0.7\pi$ corresponding to $n=n_c+n_f=1.4$.

We display the relationship between $E_0$ and $V^2$ in Fig. \ref{fig8}(a), where other parameters 
are fixed at their previously given values. One can see that the characteristic temperature $E_0$, 
which corresponds to the Kondo temperature in the impurity problem, increases
with increase in the hybridization parameter. 
Since the effect of pressure is simulated by an increase in the hybridization 
parameter between the conduction electrons and the f-electrons, $E_0$ rises
monotonously with increasing applied pressure.
Fig. \ref{fig8}(b) shows the $V^2$ dependence of the residue 
$a$ of the slave-boson Green function. This shows that $n_f$ ($= 1-a$) approaches unity with decreasing $V$.
This behavior is consistent with that of the renormalization factor $q$ 
derived from a mean-field-type approximation\cite{gut} and a Variational Monte Carlo study\cite{gut2} 
on the basis of the Gutzwiller ansatz:
%
%
\begin{align}
q^{-1}=\frac{1-n_f/2}{1-n_f}.
\end{align}
%
%
\begin{figure}[t]
\begin{center}
\includegraphics[scale=0.45]{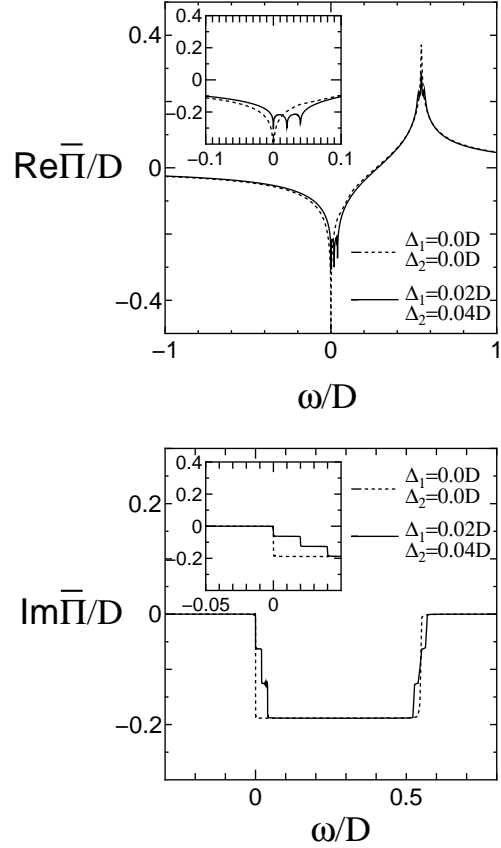}
\end{center}
\caption{Self-energy of the slave-boson $\bar{\Pi}(\omega)$ for the case of 
$\Delta_1=\Delta_2=0$, and $\Delta_1=0.02D\ \Delta_2=0.04D$.
Insets show magnified section around $\omega=0$.}
\label{fig9}
\end{figure}
%
%
%
%
\subsection{Spectral weight of slave-boson}
\begin{figure}[t]
\begin{center}
\includegraphics[scale=0.4]{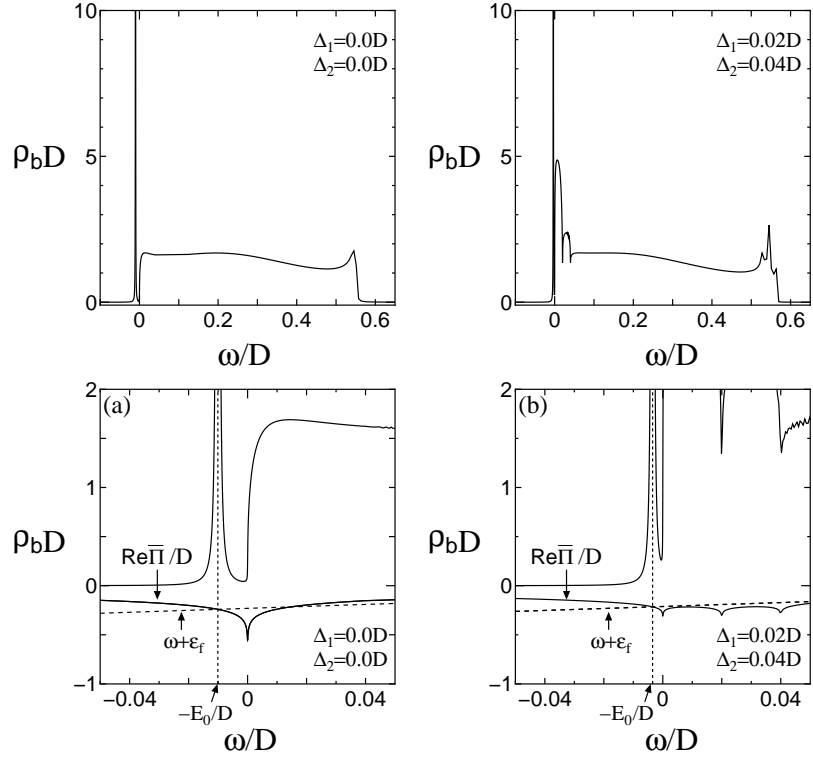}
\end{center}
\caption{Spectral function $\rho_b(\omega)$ of the slave-boson. In panels (a) and (b),
 the region $\omega \sim 0$ is enlarged, and the real part of $\bar{\Pi}(\omega)$ is also shown.}
\label{fig10}
\end{figure}

We calculate the spectral weight of the slave-boson at $T=0$. 
The spectral function $\rho_b(\omega)$ of the slave-boson is given by
%
%
\begin{align}
\rho_b(\omega)\equiv-\frac{1}{\pi} \textrm{Im} \bar{B}(\omega_{+})=\frac{1}{\pi }\cdot
\frac{0_+-\textrm{Im}\bar{\Pi}(\omega_{+})}{[ \omega+\varepsilon_f-\textrm{Re}\bar{\Pi}(\omega_{+})]^2
+[0_+- \textrm{Im}\bar{\Pi}(\omega_{+})]^2}.
\end{align}
We adopt following parameters; $D=1$, $n \equiv n_c+n_f=1.4$, $V^2=0.02D^2$, $\varepsilon_f^{(0)}=-0.7D$,
and show the results in the absence of $\Delta_1$, $\Delta_2$ and 
$\Delta_1=0.02D$, $\Delta_2=0.04D$ in Fig. \ref{fig9} and \ref{fig10}.
%
%
The function of the self-energy of the slave-boson $\bar{\Pi}(\omega)$ is 
illustrated in Fig. \ref{fig9} for two sets of the CEF level
scheme $\Delta_1=\Delta_2=0$, and $\Delta_1=0.02D$ and $\Delta_2=0.04D$. 
At $T=0$ in the absence of CEF splitting, 
$\textrm{Im}\bar{\Pi}(\omega)$ has a sharp structure at $\omega=0$ due to 
the Fermi distribution function, i.e., from eq. (2.51)
%
%
\begin{align}
\textrm{Im}\bar{\Pi}(\omega_+)=\frac{3V^2}{N_L}\sum_{\sigma}\sum_{{\bf k}_\sigma}
f(-\omega)\textrm{Im}G_{{\bf k}\sigma}(-\omega+\textrm{i}0_+),
\end{align}
where we note that $E_{\textrm{i}\Gamma}=\varepsilon_f$ in the absence of CEF splitting
in eq. (2.51).
When we take into account the CEF splitting, three-step-like structure appears,
because $\textrm{Im}\bar{\Pi}(\omega)$ is given by
%
%
\begin{align}
\textrm{Im}\bar{\Pi}(\omega_+)
=&
\frac{V^2}{N_L}\sum_{\sigma}\sum_{{\bf k}_\sigma}
\Bigr[ f(-\omega)\textrm{Im}G_{{\bf k}\sigma}(-\omega+\textrm{i}0_+)\nonumber\\
&+\sum_{l=1,2}f(-\omega+\Delta_l)\textrm{Im}
G_{{\bf k}\sigma}(-\omega+\Delta_l+\textrm{i}0_+) \Bigl].
\end{align}

Fig. \ref{fig10} shows the spectral function of the slave-boson $\rho_b(\omega)$, 
for the same sets of the CEF level scheme. One can see that the resonant peak is exhibited at
$\omega=E_0$ and a broad continuum appears in the region of $\omega >0$.
The condition, $\omega+\varepsilon_f-\textrm{Re}\bar{\Pi}(\omega)=0$, is satisfied only
when $\omega=-E_0$ $(E_0>0)$ in the region of $\omega <0$ in Fig. \ref{fig10}, while 
$\textrm{Im}\bar{\Pi}(\omega)=0$ at $\omega<0$ in Fig. \ref{fig9}.
For this reason, $\rho_b(\omega)$ has a resonant peak at $\omega=-E_0$, 
such as $1/(\omega+E_0+\textrm{i}0_+)$.
In the region of $0<\omega<\omega_{\textrm{edge}}$, where $\omega_{\textrm{edge}}$ is 
defined by a condition $\textrm{Im}G_{{\bf k}\sigma}(\omega\ge\omega_{\textrm{edge}})=0$,
$\textrm{Im}\bar{\Pi}$ has a finite value in Fig. \ref{fig9},
so that $\rho_b(\omega)$ has a broad peak in this region.
This result is consistent with eq. (2.54).
%
%
%
\subsection{Spectral weight of conduction electron and f-electron}
\begin{figure}
\begin{center}
\includegraphics[scale=0.35]{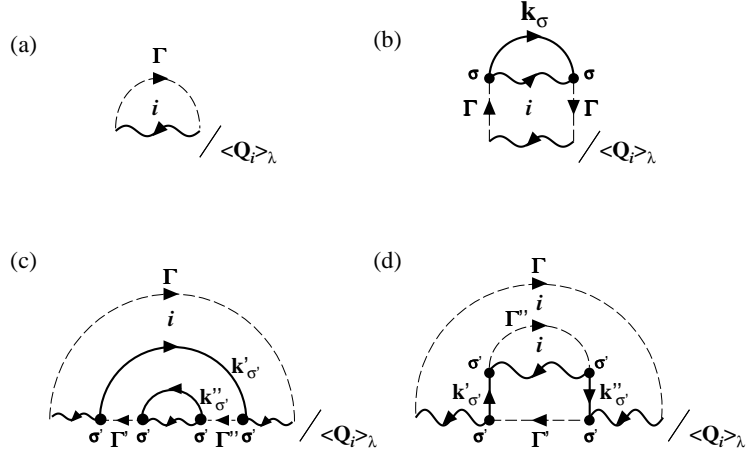}
\end{center}
\caption{Lowest order contribution to $G^{i}_{f}$ in 1/$N$-expansion.}
\label{fig11}
\end{figure}

Next, we calculate the one-particle spectral weight $\rho_c(\omega)$ of the conduction electron 
and $\rho^i_f(\omega)$ of the f-electron.
The spectral weight $\rho_c(\omega)$ is defined by
%
%
\begin{align}
\rho_c(\omega)\equiv\frac{1}{N_L}\sum_{\sigma}\sum_{{\bf k}_\sigma}\rho_{{\bf k}\sigma}(\omega)
&=\frac{1}{N_L}\sum_{\sigma}\sum_{{\bf k}_\sigma}\Bigl[ \frac{-1}{\pi}\textrm{Im}G_{{\bf k}\sigma}(\omega_{+})\Bigr],\nonumber\\
&=\frac{1}{N_L}\sum_{\sigma}\sum_{{\bf k}_\sigma}\Bigl[ \sum_{j=1}^{4}A_{{\bf k}}^{j}\cdot f(\alpha_{{\bf k}\sigma}^{j}) \Bigr],
\end{align}
and the spectral weight $\rho^i_f(\omega)$ of f-electrons is as follows:
%
%
\begin{align}
\rho^{i}_f(\omega)&\equiv-\frac{1}{\pi}\lim_{\lambda_i \rightarrow \infty}
\Bigl[  \sum_{\Gamma\pm} \textrm{Im} G^{i,\lambda}_{f\Gamma}(\omega_{+}) /\langle \hat{Q}_i \rangle_{\lambda}\Bigr],\\
&=-\frac{1}{\pi}\Bigl[  \textrm{Im} G^{i}_{f}(\omega_{+}) \Bigr],
\end{align}
where
%
%
\begin{align}
G^{i,\lambda}_{f\Gamma}(\textrm{i}\omega_n)&\equiv -\int_{0}^{\beta} \textrm{e}^{\textrm{i}\omega_n \tau}\langle \textrm{T}\tau
[f_{i\Gamma}(\tau)b_i^{\dag}(\tau)b_i f_{i\Gamma}^{\dag}]\rangle_{\lambda} \textrm{d}\tau,\\
G^{i}_{f}(\omega_+)&=\lim_{\lambda_i \rightarrow \infty}\sum_{\Gamma\pm}G^{i,\lambda}_{f\Gamma}(\omega_+)
/\langle \hat{Q}_i \rangle_{\lambda}=\sum_{\Gamma\pm}G_{f\Gamma}^{i}(\omega_+).
\end{align}
Here, it is noted that the annihilation operator $f_{\textrm{phys},i}$ of the physical f-electron at the $i$-th 
site is given by the product of the operators of the pseudo-fermion and the slave-boson as 
$f_{\textrm{phys},i}=f_{i\Gamma}b_i^{\dag}$.
Namely, the Green function $G^{i}_{f}$ for f-electron is obtained by the convolution of slave-boson 
and pseudo-fermion Green function. The lowest order contribution to $G^{i}_{f}$ in the 1/$N$-expansion 
is illustrated in Fig. \ref{fig11}. The diagrams (b), (c), (d) in Fig. \ref{fig11},
when the summations on $\sigma$ were not performed,
give higher order corrections to the self-energy of the conduction electron,
according to the classification rule of the Feynman diagrams in \S2.
However, when we calculate the spectral weight, an extra summation with respect to 
$\sigma$ must be performed, so that the terms are returned to the
lowest contributions.
%
%
The first term (a) in Fig. \ref{fig11}, which is denoted as $G^{i}_{f}(\omega_+)_{\textrm{(a)}}$, is given as follows:
%
%
\begin{align}
G^{i}_{f}(\textrm{i}\omega_n)_{\textrm{(a)}}
=-\lim_{\lambda_i \rightarrow \infty}\sum_{\Gamma\pm} T\sum_{\nu_n}B_i(\textrm{i}\nu_n)F^0_{i\Gamma}
(\textrm{i}\omega_n+\textrm{i}\nu_n)/\langle \hat{Q}_i \rangle_{\lambda},
\end{align}
where $B_i(i\nu_n)$ and $F^{0}_{i\Gamma}(i\omega_n+i\nu_n)$ are given by eqs. (2.17) and (2.18), respectively.
The convolution of ${B}_i$ and $F^0_{i\Gamma}$, eq. (3.26), 
is equal to $\Sigma_{{\bf k}\sigma}/|V_{{\bf k}\Gamma\sigma}|^2$
as can be seen in eq. (2.37) if the k-dependence of $V_{{\bf k}\Gamma\sigma}$ is neglected 
as in the present approximation
scheme. On this basis, with the use of eq. (3.6) for $\Sigma_{{\bf k}\sigma}$,
we obtain
%
%
\begin{align}
G^{i}_{f}(\omega_{+})_{\textrm{(a)}}
=\sum_{\Gamma\pm}\frac{a}{\omega_{+}-E_0-E_{i\Gamma}+\varepsilon_f}.
\end{align}
Then, a contribution to the average number from $G^{i }_{f}(\omega)_{\textrm{(a)}}$ is
%
%
\begin{align}
\int d\omega f(\omega) \Bigl[ \frac{-1}{\pi}\textrm{Im}G^{i}_{f}(\omega_{+})_{\textrm{(a)}}\Bigr] = \sum_{\Gamma\pm}a f(E_0+E_{i\Gamma}-\varepsilon_f)=0,
\end{align}
where $E_0> 0$, so that $E_0+E_{i\Gamma}-\varepsilon_f > 0$. $\textrm{Im}G^{i}_{f}(\omega_{+})_{\textrm{(a)}}$ has
a value only in the region of $\omega >0$.
Therefore, the contribution from 
 the first term (a) in Fig. \ref{fig11} vanishes. The terms (c) and (d) in Fig. \ref{fig11} also have a value
only in the region of $\omega >0$, so that they do not contribute to the average number of f-electrons.
As a result only the second term (b) in Fig. \ref{fig11} has a value in the region of $\omega<0$ and
contributes to the average number of f-electrons at $T=0$. For this reason, when we discuss the spectral weight
of the f-electron in the region of $\omega<0$ at $T=0$, we should only take into account term (b).
The analytical form of the second term (b) at $T=0$ is given by (See Appendix A for details)
%
%
\begin{align}
G^{i}_{f}(\omega_{+})_{\textrm{(b)}}=&\frac{1}{N_L}\sum_{\Gamma\pm}\sum_{\sigma}\sum_{{\bf k}_\sigma}|V_{\Gamma \sigma}|^2
\int \textrm{d} \varepsilon f(\varepsilon) \Bigl[ \frac{-1}{\pi}\textrm{Im}
 G_{{\bf k}\sigma} (-\varepsilon+\textrm{i}0_+)\Bigr] \int \textrm{d} \varepsilon' \textrm{e}^{-\beta (E_0+\varepsilon')}\nonumber\\
&\times\Bigl[ \frac{-1}{\pi}\textrm{Im}\bar{B}(\varepsilon'+\textrm{i}0_+)\Bigr]
\bar{B}(\omega_{+} +\varepsilon+\varepsilon')
\frac{1}{(\omega_{+} +\varepsilon'-E_{i\Gamma}+\varepsilon_f)^2}/\langle \hat{Q} \rangle\nonumber\\
&+\frac{-1}{N_L}\sum_{\Gamma\pm}\sum_{\sigma}\sum_{{\bf k}_\sigma}|V_{\Gamma \sigma}|^2
\int \textrm{d} \varepsilon f(\varepsilon) \Bigl[ \frac{-1}{\pi}\textrm{Im}
 G_{{\bf k}\sigma} (\varepsilon+\textrm{i}0_+)\Bigr] \int \textrm{d} \varepsilon' \textrm{e}^{-\beta (E_0+\varepsilon')}\nonumber\\
&\times\Bigl[ \frac{-1}{\pi}\textrm{Im}\bar{B}(\varepsilon'+\textrm{i}0_+)\Bigr]
\bar{B}(-\omega+\varepsilon+\varepsilon'+\textrm{i}0_+)
\frac{1}{(\varepsilon +\varepsilon'-E_{i\Gamma}+\varepsilon_f)^2}/\langle \hat{Q} \rangle,
\end{align}
where we have used the previous notation: $\langle \hat{Q}_i \rangle_{\lambda}=\textrm{e}^{-\beta(\lambda_i+\varepsilon_f-E_0)}\langle \hat{Q} \rangle$.
Substituting this into eq. (3.23), we obtain
%
%
\begin{align}
\rho^{i}_f(\omega)=&\frac{1}{N_L}\sum_{\Gamma\pm}\sum_{\sigma}\sum_{{\bf k}_\sigma}|V_{\Gamma \sigma}|^2
\int \textrm{d} \varepsilon \ f(\varepsilon) \int \textrm{d}\varepsilon'
\textrm{e}^{-\beta(E_0+\varepsilon')}\frac{-1}{\pi}\textrm{Im}\bar{B}(\varepsilon '+\textrm{i}0_+)/\langle \hat{Q}\rangle\nonumber\\
&\times \Bigl[ \frac{-1}{\pi}\textrm{Im}G_{{\bf k}\sigma}(-\varepsilon+\textrm{i}0_+)\cdot \frac{-1}{\pi}
\textrm{Im}\bar{B}(\omega+\varepsilon+\varepsilon'+\textrm{i}0_+)
\cdot \frac{1}{(\omega+\varepsilon'-E_{i\Gamma}+\varepsilon_f)^2}\nonumber\\
&+\frac{-1}{\pi}\textrm{Im}G_{{\bf k}\sigma}(\varepsilon+\textrm{i}0_+)\cdot \frac{-1}{\pi}
\textrm{Im}\bar{B}(-\omega+\varepsilon+\varepsilon'+\textrm{i}0_+)\cdot
\frac{1}{(\varepsilon+\varepsilon'-E_{i\Gamma}+\varepsilon_f)^2} \Bigr].
\end{align}
With the use of (2.54) at $T=0$, this is reduced to
%
%
\begin{figure}[t]
\begin{center}
\includegraphics[scale=0.4]{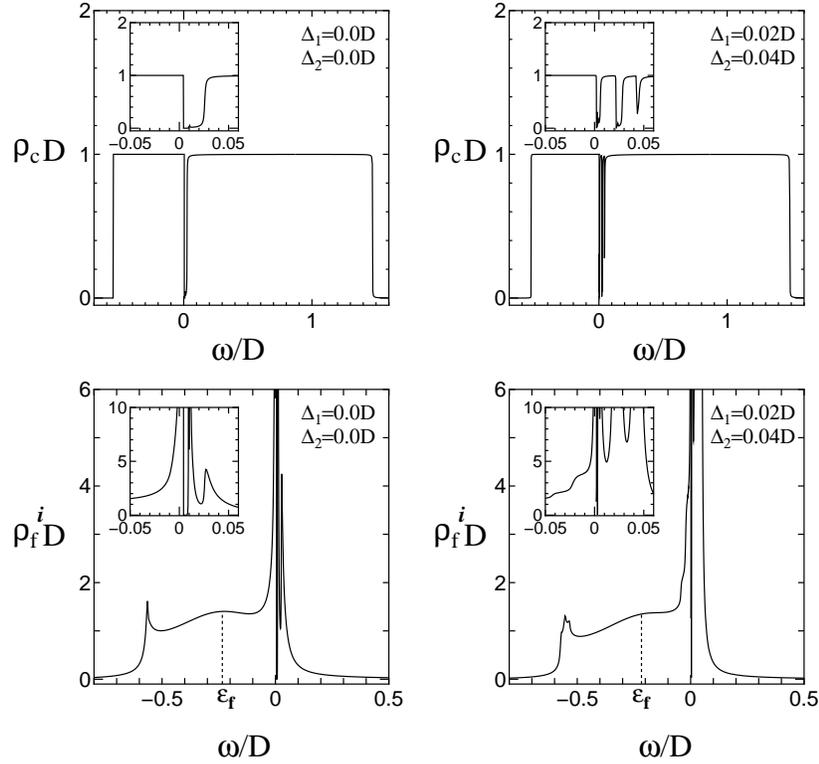}
\end{center}
\caption{Spectral weight of the conduction electron and f-electron.
Insets are magnified views around $\omega=0$. }
\label{fig12}
\end{figure}
%
%
\begin{align}
\rho^{i}_f(\omega)=& \sum_{\Gamma\pm}\sum_{\sigma} \Bigl( \frac{a |V_{\Gamma \sigma}|}{\omega -E_0-E_{i\Gamma}+\varepsilon_f}\Bigr)^2 \rho_\sigma(\omega)
\\
&+  \sum_{\Gamma\pm}\sum_{\sigma}\frac{a|V_{\Gamma \sigma}|^2}{(\omega -E_0-E_{i\Gamma}+\varepsilon_f)^2} 
\int \textrm{d} \varepsilon \ f(\varepsilon) 
\rho_\sigma(-\varepsilon) C(\omega+\varepsilon-E_0)\\
&+ \sum_{\Gamma\pm}\sum_{\sigma}\int \textrm{d} \varepsilon \ f(\varepsilon)
\rho_\sigma(\varepsilon) C(-\omega+\varepsilon-E_0)\frac{a|V_{\Gamma \sigma}|^2}{(\varepsilon-E_0-E_{i\Gamma}+\varepsilon_f)^2}.
\end{align}
Here, $\rho_\sigma$ is defined in eq. (3.15), and $C(\omega)$ is the broad continuous function in eq. (2.54).
The contribution from eq. (3.31) corresponds to the coherent part of the spectral weight of f-electrons, 
and that from eqs. (3.32) and (3.33) give its incoherent part. 

We show the numerical results of $\rho_c(\omega)$, eq. (3.21), and $\rho_f^i(\omega)$, eqs. (3.31)$\sim$(3.33),
in Fig. \ref{fig12}, where we adopt the following parameters: $D=1$, 
$n \equiv n_c+n_f=1.4$, $V^2=0.02D^2$, $\varepsilon^{(0)}_f=-0.7D$.
A resonance-like peak is observed at $\omega=E_0$ in $\rho^i_f(\omega)$, the spectrum of the f-electron.
This peak arises from the term, eq. (3.32), in which $C(\omega)$ is a broad function of $\rho_b(\omega)$ in the region of $\omega >0$.
Physically speaking, it reflects a excitation process from the binding state of the slave-boson
to continuum states whose minimum excitation energy is given by $E_0$.
This result can not be obtained by a simple quasi-particle picture of the Fermi liquid, 
where the renormalization factor $z$ is estimated as $z=(1-\partial \Sigma(\omega)/\partial \omega|_{\omega=0})^{-1}$,
but is related to a subtle structure of incoherent states with excitation energy of the order $E_0\sim T_F^*$,
$T_F^*$ being a renormalized Fermi energy. 
Here, with the use of these spectral weight functions, we find that the following sum-rule holds within an error of a few percentage at worse.
%
%
\begin{eqnarray}
\int_{-\infty}^{\infty} \textrm{d} \omega \   f(\omega) \Bigl[ \rho_{c}(\omega)+ \rho_f^{i}(\omega) \Bigr] =n.
\end{eqnarray}
%
%
%
\subsection{Kondo temperature}
$E_0$ defined as the binding energy of the slave-boson at $T=0$ corresponds to the Kondo temperature $T_K$
defined in an impurity version of the Anderson model. In the impurity problem, the conduction electron is 
not renormalized, so that we can put  $E_0 \rightarrow T_K$, $\alpha_{{\bf k}\sigma}^{j}\rightarrow \varepsilon_{{\bf k}\sigma}$
and $A_{\bf k}^{j}\rightarrow 1$. Then, we only have to take into account eq. (3.12), leading to
%
%
\begin{align}
T_K=&D\cdot(\frac{D}{\Delta_1})\cdot(\frac{D}{\Delta_2})\exp\Bigl(
-\frac{|\varepsilon_f|}{2\rho_0V^2} \Bigr).
\end{align}
Similarly, we can calculate the Kondo temperature $T_K^{(0)}$ in the absence of CEF splitting.
Then, we can derive the following relation between $T_K$ and $T_K^{(0)}$ under the condition
$T_K \ll \Delta_1\sim\Delta_2 \ll |\varepsilon_f| < |D|$,
%
%
\begin{figure}[t]
\begin{center}
\includegraphics[scale=0.45]{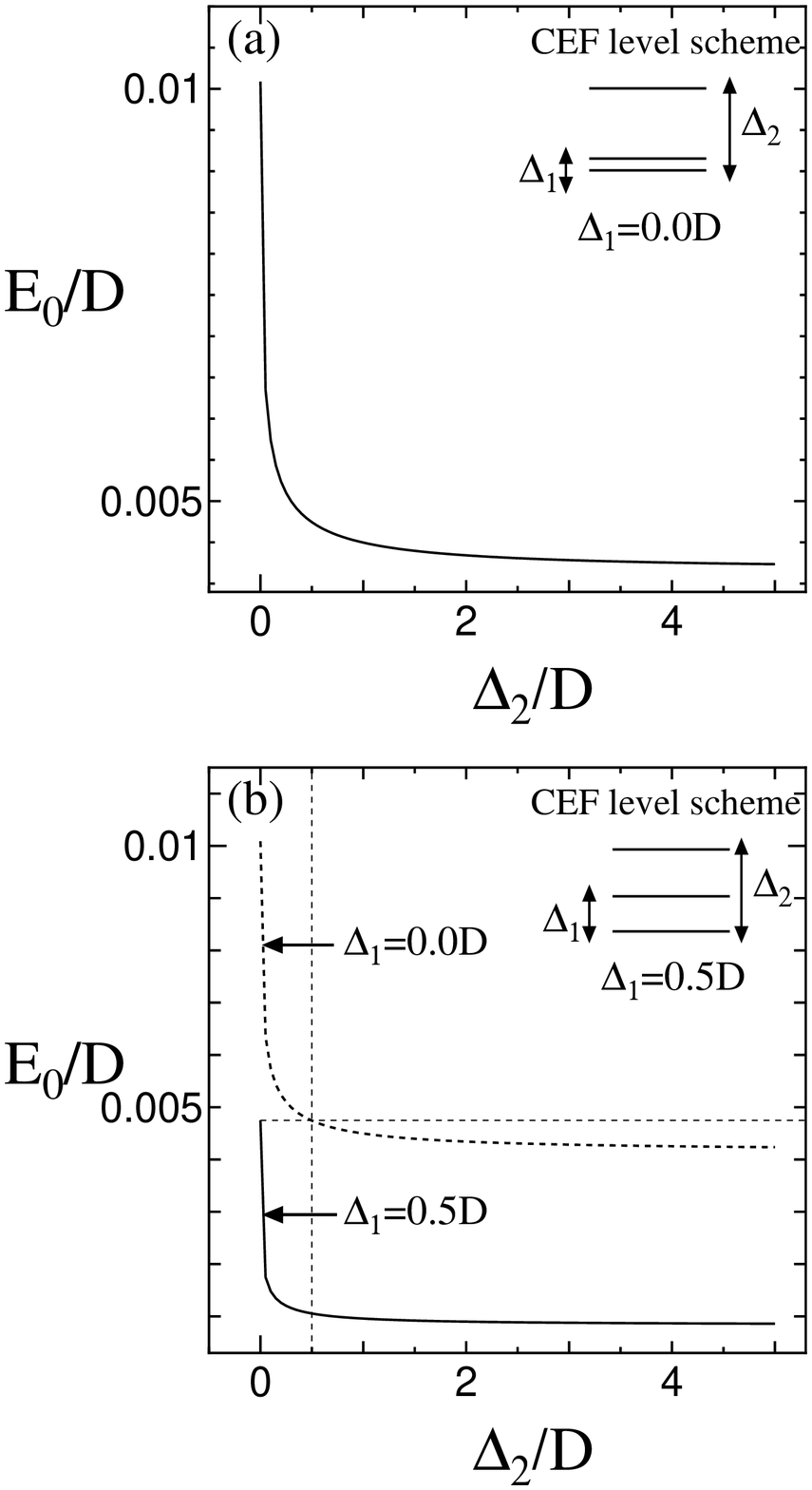}
\end{center}
\caption{
(a) $\Delta_2$ vs $E_0$  in the case of $\Delta_1=0$.
(b) $\Delta_2$ vs $E_0$  in the both case of $\Delta_1=0$ (dot line) 
and $\Delta_1=0.5D$ (solid line).}
\label{fig13}
\end{figure}
%
%
\begin{eqnarray}
\Bigl[ T_K^{(0)} \Bigr]^3= \Delta_1\cdot\Delta_2\cdot T_K.
\end{eqnarray}
(See Appendix B for derivation) 
This result is consistent with previous works on the impurity problem \cite{yama,han}.

Next, we consider the lattice problem, for which we must solve coupled self-consistent 
equations (3.12)$\sim$(3.14).
By solving these equations,
we can evaluate $E_0$, which is regarded as the Kondo temperature in the lattice case. 
We illustrate the numerical results for the effect of CEF splitting on $E_0$ in Fig. \ref{fig13}, 
where we adopt the following parameters; $D=1$, $n \equiv n_c+n_f=1.4$, $V^2=0.02D^2$, 
$\varepsilon^{(0)}_f=-0.7D$, and (a) $\Delta_1=0.0D$, and (b) $\Delta_1=0.5D$.

In case (a), if $\Delta_2=0.0D$, $E_0$ should be in agreement with $E_0^{(0)}$,
which is defined as the Kondo temperature in the absence of CEF splitting in the lattice case.
Here, with the use of the relations (3.35) and (3.36) derived in the impurity case, we can roughly estimate
$E_0^{(0)}$ from $T_{K}^{(0)}$.
Substituting the present set of parameters: $\rho_0 =0.5/D$, $\varepsilon_f\equiv\varepsilon_f^{(0)}-\mu=-0.23D$, 
and $V^2=0.02D^2$, we obtain
%
%
\begin{eqnarray}
T_K^{(0)}  = D\exp\Bigl(-\frac{0.23}{6 \times 0.5 \times 0.02} \Bigr)\simeq 2.16\times10^{-2}D.
\end{eqnarray}
Numerical results in Fig. \ref{fig13}(a) show $E_0^{(0)}\sim 10^{-2}D$, which is reasonable compared with the 
rough estimate of eq. (3.37).

One can see in Fig. \ref{fig13}(a) that 
$E_0$ decreases rapidly with increasing $\Delta_2$, and $E_0$ saturates at $\Delta_2 \simeq 4D$.
This shows that the Kondo temperature shifts from that with 6-fold degeneracy to that with 4-fold 
degeneracy. Such a behavior was not derived from the SU($N$)-PAM+CEF model, in which $E_0$ becomes a 
negative value at large $\Delta$ $\sim$ $4D$.
Namely, the application of the $1/N$-expansion method to SU($N$)-PAM fails when we take into account the CEF splitting.

In case (b), $E_0$ also decreases rapidly with increasing $\Delta_2$
as shown in Fig. \ref{fig13}(b). This also shows that
the Kondo temperature shifts from that with 4-fold degeneracy to that with 2-fold degeneracy.
We find that the value of $E_0\sim4.7\times10^{-3}D$ at $\Delta_1=0.0D$, $\Delta_2=0.5D$ is 
consistent with that of $E_0$ at $\Delta_1=0.5D$, $\Delta_2=0.0D$.
%
%
%
\section{Physical properties at Finite Temperatures}
%
%
\begin{figure}[t]
\begin{center}
\includegraphics[scale=0.45]{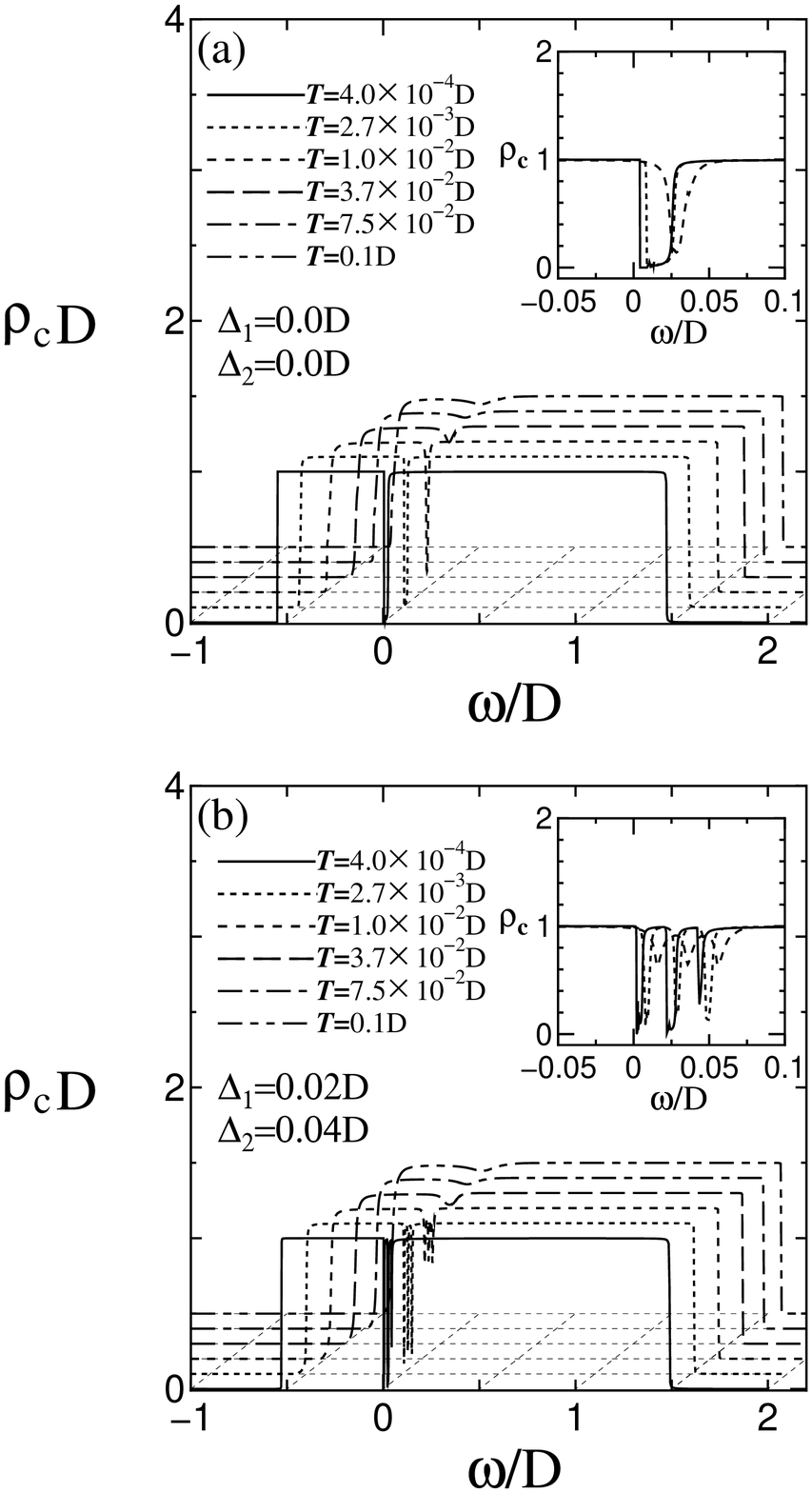}
\end{center}
\caption{Spectral weight of conduction electron at each temperature 
for cases (a) $\Delta_1=\Delta_2=0.0D$ and (b) $\Delta_1=0.02D$ $\Delta_2=0.04D$. 
Insets are magnified views around $\omega=0$.}
\label{fig14}
\end{figure}
In this section, we discuss the resistivity over the entire temperature range.
To this end, we perform the calculations beyond the low-temperature 
approximation given in \S 3. In other words, when we discuss the physical properties at high temperatures ($T \gg E_0$), 
it is not appropriate to use eq. (2.54) given in $\S3$, since $\textrm{Im}\bar{\Pi}$ has the width of 
$\sim T$ near $\omega=0$ due to the thermal smearing effect of the Fermi distribution function. Namely, 
the picture of the resonant peak at $E_0$ ceases to be valid so that we must take into account the entire 
structure of the spectrum of the slave-boson and the contribution from the pole of pseudo-fermion at high temperatures. 
Then, we must solve the Dyson equations eqs. (2.16)$\sim$(2.18) for the single-particle Green functions self-consistently for all values of frequency $\omega$ at each temperature.
Details of the self-consistent calculation are discussed in Appendix C.
%
%
%
\subsection{Spectral weight of conduction electron}
The spectral weight of $\rho_c(\omega)$ of the conduction electron, given by eq. (3.21), is shown in Fig. {\ref{fig14}} for a series of temperatures.
The parameters adopted are the same as for Fig. {\ref{fig14}}$\sim$Fig. {\ref{fig19}} : $D=1$, $n \equiv n_c+n_f=1.4$, $V^2=0.02D^2$, $\varepsilon^{(0)}_f=-0.7D$.
Fig. {\ref{fig14}}(a) is for the case without CEF splitting ($\Delta_1=\Delta_2=0$), and Fig. {\ref{fig14}}(b)
is for the case with CEF splitting $\Delta_1=0.02D$ and $\Delta_2=0.04D$.
Here, we have also assumed the DOS of bare conduction electron per spin as follows:
%
%
\begin{align}
\rho_{\sigma}(\omega)=\frac{1}{N_L}\sum_{{\bf k}_\sigma}\rho_{{\bf k}\sigma }= \left\{
                   \begin{array}{@{\,}ll}
                   \rho_0=\frac{1}{2D} & \mbox{($-D \le \omega \le  D$)},\\
                   0      & \mbox{(otherwise)}.
                   \end{array}
                 \right.
\end{align}
With increasing temperature, the hybridization gap in the 
spectral weight of the conduction electron is gradually buried through pseudo-gap-like behavior.
Such behavior has already been pointed out in a previous work detailed in ref. \citen{Ono2}.
It is noted that, in the case of $\Delta_1=0.02D$ and $\Delta_2=0.04D$,
three hybridization gaps, corresponding to each CEF level, appear in the low-temperature limit. 
%
%
\begin{figure}[t]
\begin{center}
\includegraphics[scale=0.4]{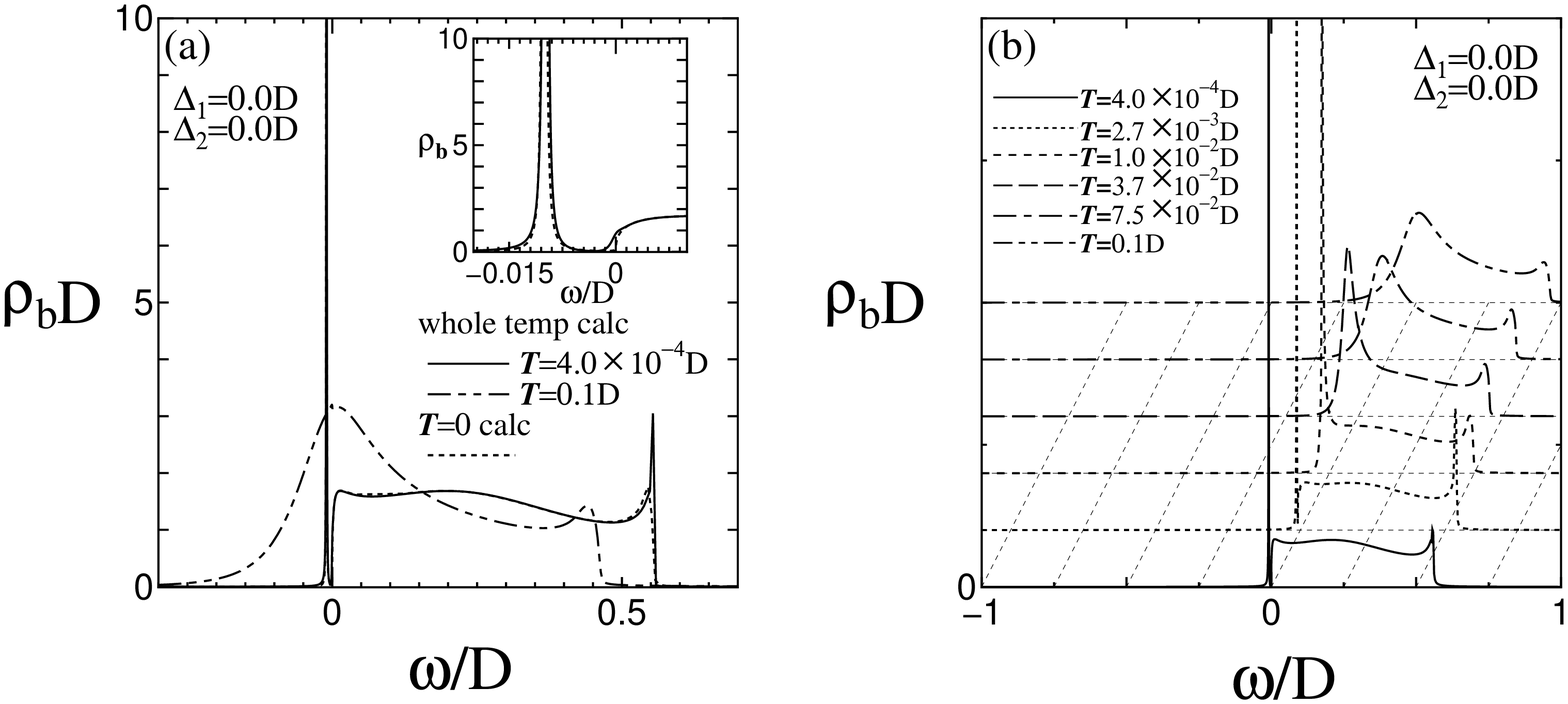}
\end{center}
\caption{(a) Spectral weight of slave-boson $\rho_b(\omega)$. Inset shows magnified view around $\omega=0$.
(b) $\rho_b(\omega)$ for a series of temperatures.}
\label{fig15}
\end{figure}
%
%
%
%
%
\subsection{Spectral weight of slave boson}
%
%
\begin{figure}[t]
\begin{center}
\includegraphics[scale=0.45]{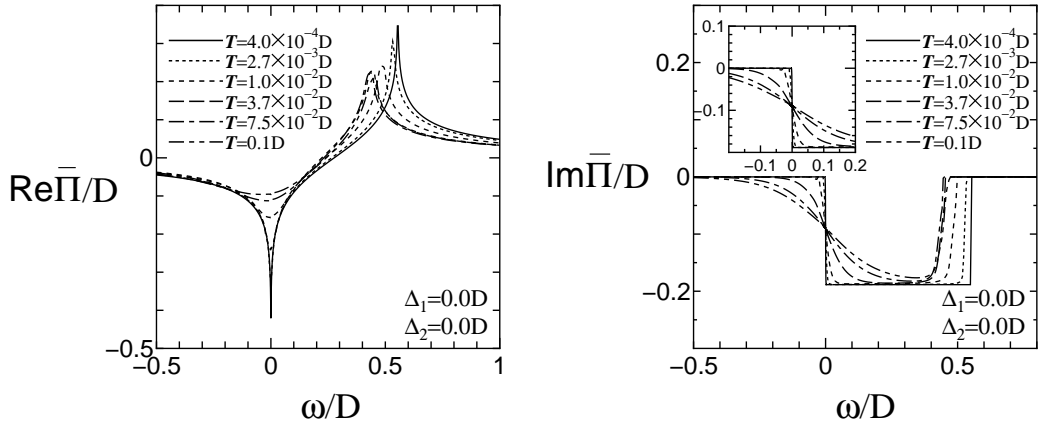}
\end{center}
\caption{Self-energy of slave-boson,
$\textrm{Re}\bar{\Pi}(\omega)$ and  $\textrm{Im}\bar{\Pi}(\omega)$,
for a series of temperatures. Inset is magnified view around $\omega=0$.}
\label{fig16}
\end{figure}
In Fig. \ref{fig15}, we show the spectral function of the slave-boson in the case without CEF splitting, i.e.,
$\Delta_1=\Delta_2=0.0D$.
In the finite temperature calculation, 
$E_0(T)$ is determined by the peak position of the spectral function of the slave-boson.
In Fig. \ref{fig15}(a), we find that $E_0(4\times10^{-4}D)$ is almost the same as the result 
$E_0(0)$ derived from the $T=0$ calculation in \S 3, $E_0(0) \simeq 0.01$. 
Namely, we can confirm that our calculation is performed from a low temperature $(T \lesssim T_0)$ 
to a high temperature $(T > E_0)$ continuously. Furthermore, we find that at a high temperature 
such as $T=0.1D$, the spectral weight of the slave-boson cannot be divided into
the resonant peak and broad function like eq. (2.54).

In Fig. \ref{fig15}(b), we show the results for a series of temperatures $4\times 10^{-4} < T/D<0.1$.
We find that the validity of eq. (2.54) collapses  at $T\sim 0.01 \sim E_0(0)$.
Indeed, the low-temperature approximation fails at temperatures higher than $E_0(0)/10$.
In the case with CEF-splitting, such a behavior does not change essentially. 
However, $E_0(0)$ becomes smaller compared with the case of $\Delta_1=\Delta_2=0.0D$ (See Fig. \ref{fig8}(a)),
and a three-step-like structure appears due to the CEF-splitting (See Fig. \ref{fig9} and \ref{fig10}).

Next, we show the result for the self-energy of the slave-boson in Fig. \ref{fig16}.
$\textrm{Im}\bar{\Pi}(\omega)$ has a sharp structure around $\omega=0$ at $T=4\times 10^{-4}D$,
while its width broadens with increasing temperature due to the smearing effect of the 
Fermi distribution function.
Therefore, $\textrm{Im}\bar{\Pi}(\omega)$ at high temperatures has a finite value 
even in the region of $\omega<0$. Hence, the peak at $\omega=E_0$ has a finite width, 
so that $\rho_b(\omega)$ cannot be divided into two parts, a resonant peak 
(in the region of $\omega<0$) and a broad background (in the region of $\omega>0$), at high temperature $T\sim E_0$.
%
%
%
\subsection{Electrical resistivity}
In this subsection, the temperature dependence of the resistivity is discussed.
We calculate the conductivity by means of the Kubo formula\cite{kubo}.
Since the dispersion of f-electrons as well as the $k$-dependence of $V$ is neglected,
the conductivity is expressed by the Kubo formula
%
%
\begin{align}
\sigma=\frac{e^2}{N_L^2} \sum_{\sigma}\sum_{\sigma'}\sum_{{\bf k}_\sigma} \sum_{{\bf k'}_\sigma}v_{{\bf k}}v_{{\bf k'}}
\lim_{q \rightarrow 0}\lim_{\omega \rightarrow 0} \frac{\textrm{Im}K^{\bf{k}\bf{k}'}_{q}(\omega_{+})}{\omega},
\end{align}
where ${ v}_{\bf k}=\nabla_{{\bf k}}\varepsilon_{{\bf k}}$, $K$ is a two-particle Green function defined as
%
%
\begin{align}
K_{q}^{{\bf k}{\bf k}'}(\omega_{+})=\textrm{i} \int_{0}^{\infty}
\textrm{d}t\ e^{\textrm{i}\omega_{+}t} \langle [ c^{\dag}_{{\bf k}\sigma}(t)
 c_{{\bf k}+{\bf q}\sigma}(t), c^{\dag}_{{\bf k'}\sigma'}(0) c_{{\bf k'-{\bf q}}\sigma'}(0)]\rangle.
\end{align}
Since we are interested in the qualitative behavior of the resistivity, 
we neglect the vertex correction corresponding to the inverse collision process in the Landau-Boltzmann
equation. However, the correction of the current vertex is taken into account through the Ward-Pitaevskii
identity in the k-limit although this gives no net effect to the current vertex provided that the wave-number dependence of the
self-energy $\Sigma_{f}({\bf k},\omega)$ of the f-electron can be neglected compared with its frequency dependence.
Indeed, the Ward-Pitaevskii identity for the current vertex\cite{agd} is given by
%
%
\begin{align}
-\frac{{\bf p}}{m}+\frac{\textrm{i}}{2}\int \Gamma^{{\bf k}}_{\alpha\beta,\alpha\beta}(p,q)\frac{{\bf q}}{m}\{G^2(q) \}_{\bf k}
\frac{\textrm{d}^4 q}{(2\pi)^4}
=-\frac{{\bf p}}{m^* a'},
\end{align}
where $a'$ is the quasi-particle weight, and $G$ is a one-particle Green function derived from the relation between
the two-particle Green function and the vertex part.
The right-hand side of eq. (4.4) can be approximated by $-{\bf p}/m$ if the k-dependence of $\Sigma_{f}$ is neglected.
By assuming implicitly that the Umklapp process works to violate the conservation of lattice momentum, the effect of collision
is taken into account only through the self energy of $G_{{\bf k}\sigma}$.
%
%
\begin{figure}[t]
\begin{center}
\includegraphics[scale=0.4]{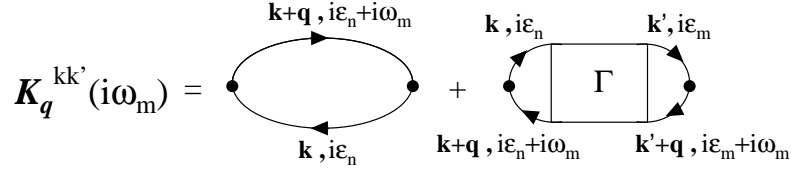}
\end{center}
\caption{Diagrammatic representation of the current-current response function.\\}
\label{fig17}
\end{figure}
Then, $\textrm{Im}K$ is given as (See Fig. \ref{fig17})
%
%
\begin{align}
\textrm{Im}K_{ q}(\omega_{+})=&-\delta_{{\bf k}{\bf k'}}\int \textrm{d} \varepsilon\ 
f(\varepsilon)\Bigl[ \frac{-1}{\pi}\textrm{Im}G_{{\bf k}\sigma}(\varepsilon+\textrm{i}0_+) \Bigr]\nonumber\\
&\times\textrm{Im} \Bigl\{ G_{{\bf k}+{\bf q}\sigma}(\varepsilon+\omega+\textrm{i}0_+)-G_{{\bf k}+{\bf q}\sigma}(\varepsilon-\omega+\textrm{i}0_+) \Bigr\}.
\end{align}
Substituting this into eq. (4.2) and performing some rearrangements, we obtain
%
%
\begin{align}
\sigma & =
\frac{e^2}{\pi N_L}\sum_{\sigma} \sum_{{\bf k}_\sigma} v_{\bf k}^2\int \textrm{d}\varepsilon \ \Bigl( -\frac{\partial f(\varepsilon)}
{\partial \varepsilon} \Bigr)\Bigl[ \textrm{Im} G_{{\bf k}\sigma}(\varepsilon+\textrm{i}0_+)\Bigr]^2.
\end{align}
%
%
\begin{figure}[t]
\begin{center}
\includegraphics[scale=0.4]{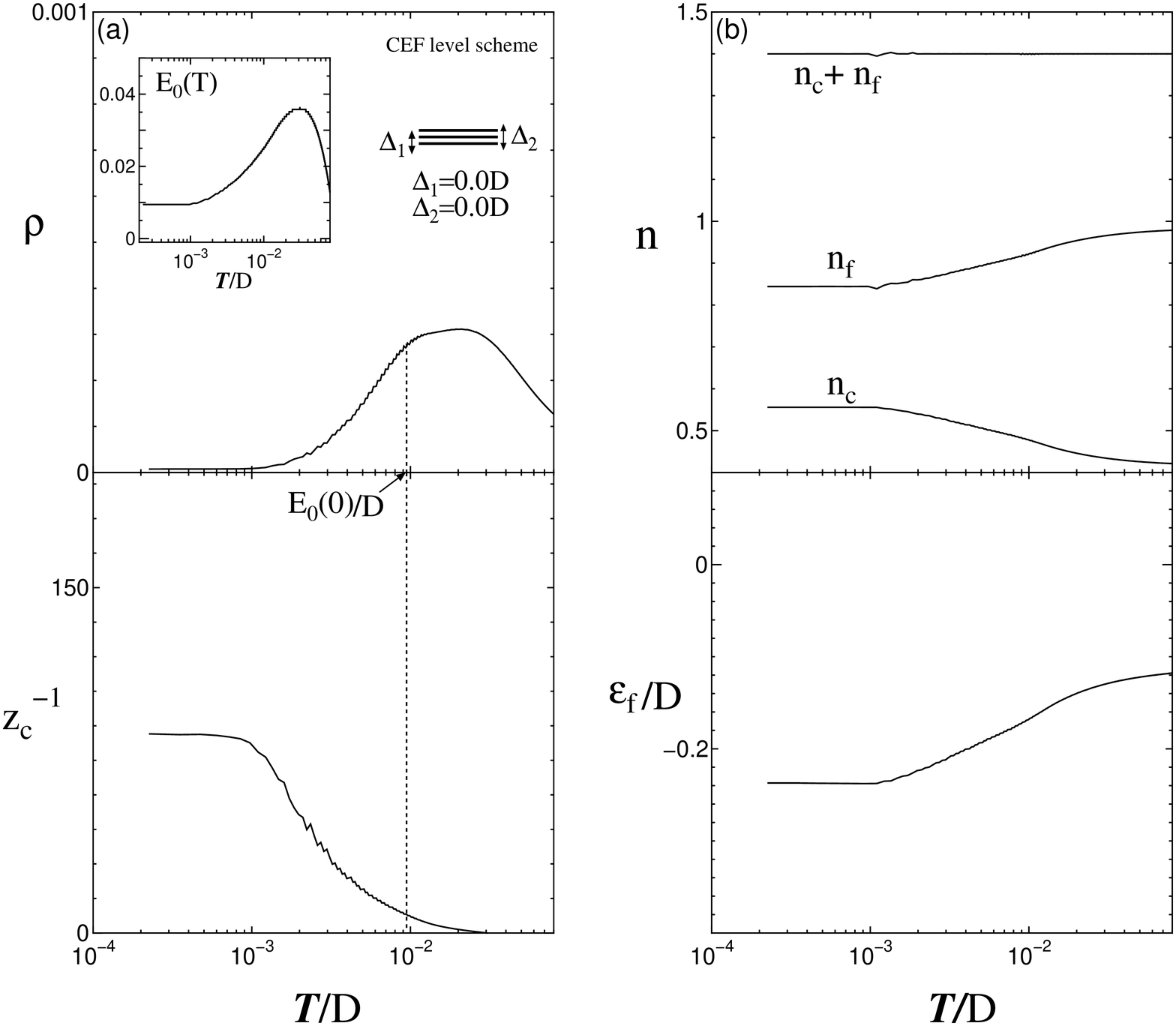}
\end{center}
\caption{(a) Temperature dependence of electrical resistivity $\rho(T)$ and $z_{c}^{-1}$.
Inset is temperature dependence of $E_0$.
(b) Temperature dependences of number of conduction electrons $n_c$, f-electron $n_f$, and
$n=n_c+n_f$, and $\varepsilon_f$ measured from the chemical potential at a finite temperature.}
\label{fig18}
\end{figure} 
%
%
By omitting some constant factors, we define the reduced conductivity,
\begin{align}
\tilde{\sigma} & \equiv \frac{1}{N_L}
\sum_{\sigma}\sum_{{\bf k}_\sigma} v_{\bf k}^2\int \textrm{d}\varepsilon \ \Bigl( -\frac{\partial f(\varepsilon)}
{\partial \varepsilon} \Bigr)\Bigl[ \textrm{Im} G_{{\bf k}\sigma}(\varepsilon+\textrm{i}0_+)\Bigr]^2.
\end{align}
With the use of the renormalized Green function for $\textrm{Im}G_{{\bf k}\sigma}$ derived from \S 4.1,
$\tilde{\sigma}$ is calculated numerically, and the reduced resistivity is defined as
$\rho={\tilde{\sigma}}^{-1}$.
 A result of the resistivity $\rho$
is displayed in Fig. \ref{fig18} and \ref{fig19}. In the absence of CEF splitting,
the resistivity has a broad peak near $T\simeq 2E_0(0)$ as shown in Fig. \ref{fig18}(a). 
In the high-temperature region, $T>2E_0(0)$,
the resistivity shows $-\textrm{log}T$ behavior, typical of the Kondo effect.
Here, we define the mass enhancement factor $z_c$ as follows:
%
%
\begin{align}
z_{c}^{-1}= 1- \frac{\textrm{\textrm{d}}\Sigma_{{\bf k}\sigma}{(\omega)}}{\textrm{d}\omega}\Bigr|_{\omega=0_{+}},
\end{align}
which is relate to the quasi-particle weight 
$z_{\Gamma}=\Bigr( 1-\partial\Sigma_{i\Gamma}^{f}(\omega)/\partial\omega \Bigr|_{\omega=0_+}\Bigr)^{-1}$ as follows:
%
%
\begin{align}
z_{c}^{-1}&
=1+\sum_{\Gamma\pm} |V_{{\bf k}\Gamma\sigma}|^2 G_{f\Gamma}^{i}(0)^2{z_{\Gamma}}^{-1},
\end{align}
where $G_{f\Gamma}^{i}(\omega_+)$ is defined in eq. (3.25).
With decreasing temperature, 
${z_c}^{-1}$ begins to increase at around $T=E_0$ and saturates at a temperature lower than $T_0 \sim E_0/10$.
Namely, the quasi-particle is formed at a temperature lower than $T_0/10$. 
The resistivity $\rho$ exhibits a gradual decrease in this temperature region.
%
%
%
\begin{figure}[t]
\begin{center}
\includegraphics[scale=0.4]{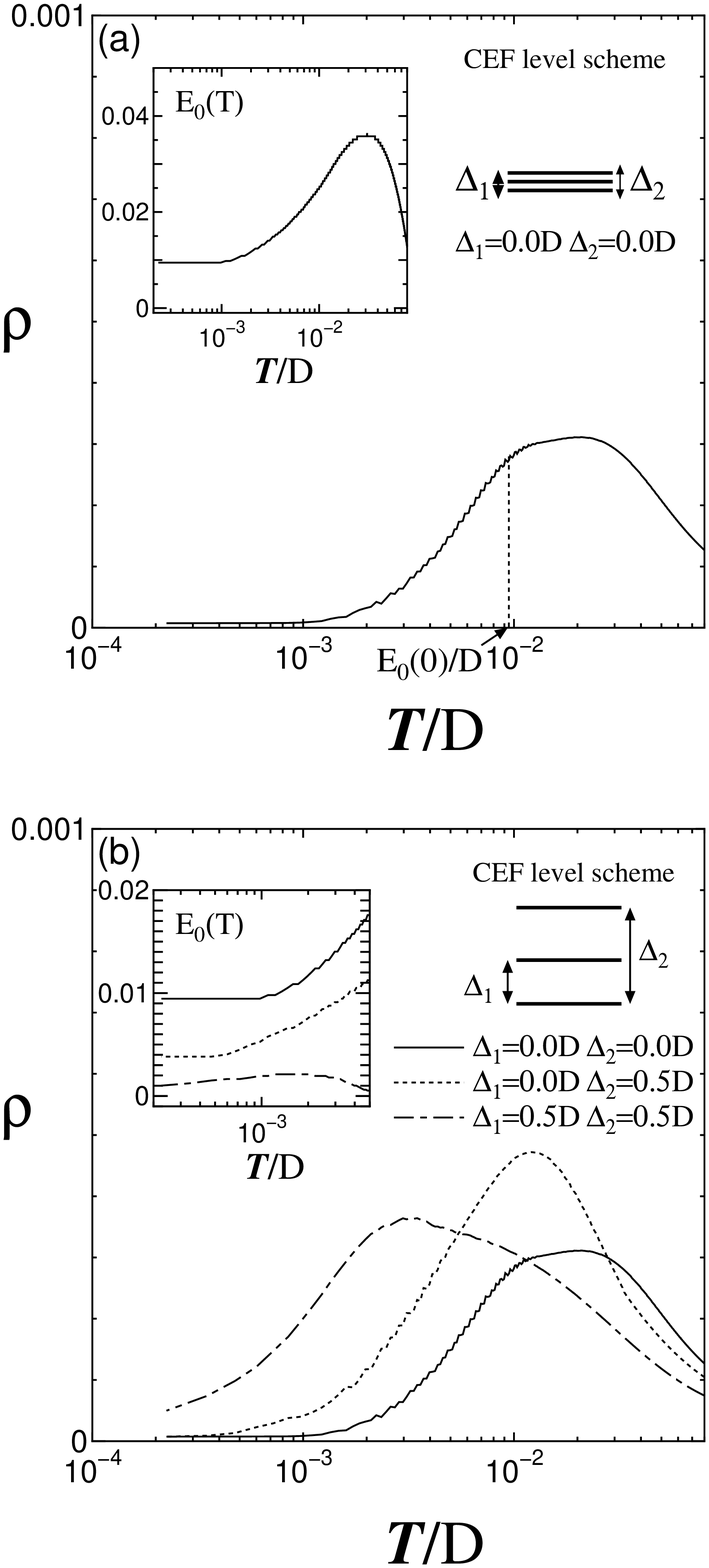}
\end{center}
\caption{(a) Temperature dependence of electrical resistivity $\rho(T)$ in case of $\Delta_1=\Delta_2=0$,
and (b) that for series of CEF-splitting schemes.}
\label{fig19}
\end{figure}
In the present calculation of the leading order in $1/N$, in which the inter-site correlation of f-electrons
cannot be taken into account, the Fermi liquid behavior $\rho \propto T^2$(at $ T\ll E_0/10$)
is not reproduced. In order to reproduce the $T^2$-behavior, we need to take into account
higher order terms of $O(1/N)^1$ as shown in ref. \citen{tsuru}.
However, the higher order correction is known not to appreciably affect
the temperature dependence of $\rho$ at $T>T_0$\cite{tsuru2}.
In Fig. \ref{fig18}(b), the $T$-dependences of $n_f$, $n_c$, and $n=n_c+n_f$,
and $\varepsilon_f$ are also displayed.
The numerical calculation is performed under the condition that $n$ is fixed.

In Fig. \ref{fig19}, the $T$-dependence of the resistivity $\rho(T)$ in both the cases of with and 
without CEF splitting are shown.
$E_0(0)$ with CEF splitting becomes smaller than that in the absence of CEF splitting.
In the inset of Fig. \ref{fig18}(a) and \ref{fig19}, the $T$-dependence of $E_0(T)$ is shown.
We find that $E_0(4\times10^{-4}D)$  $\sim0.01D$ in the case of $\Delta_1=\Delta_2=0$. 
This value is consistent with $E_0(0)$ derived from the calculation at $T=0$ in \S 3.
%
%
%
%
\section{Pressure Effect on Resistivity}
%
%
\begin{figure}[t]
\begin{center}
\includegraphics[scale=0.4]{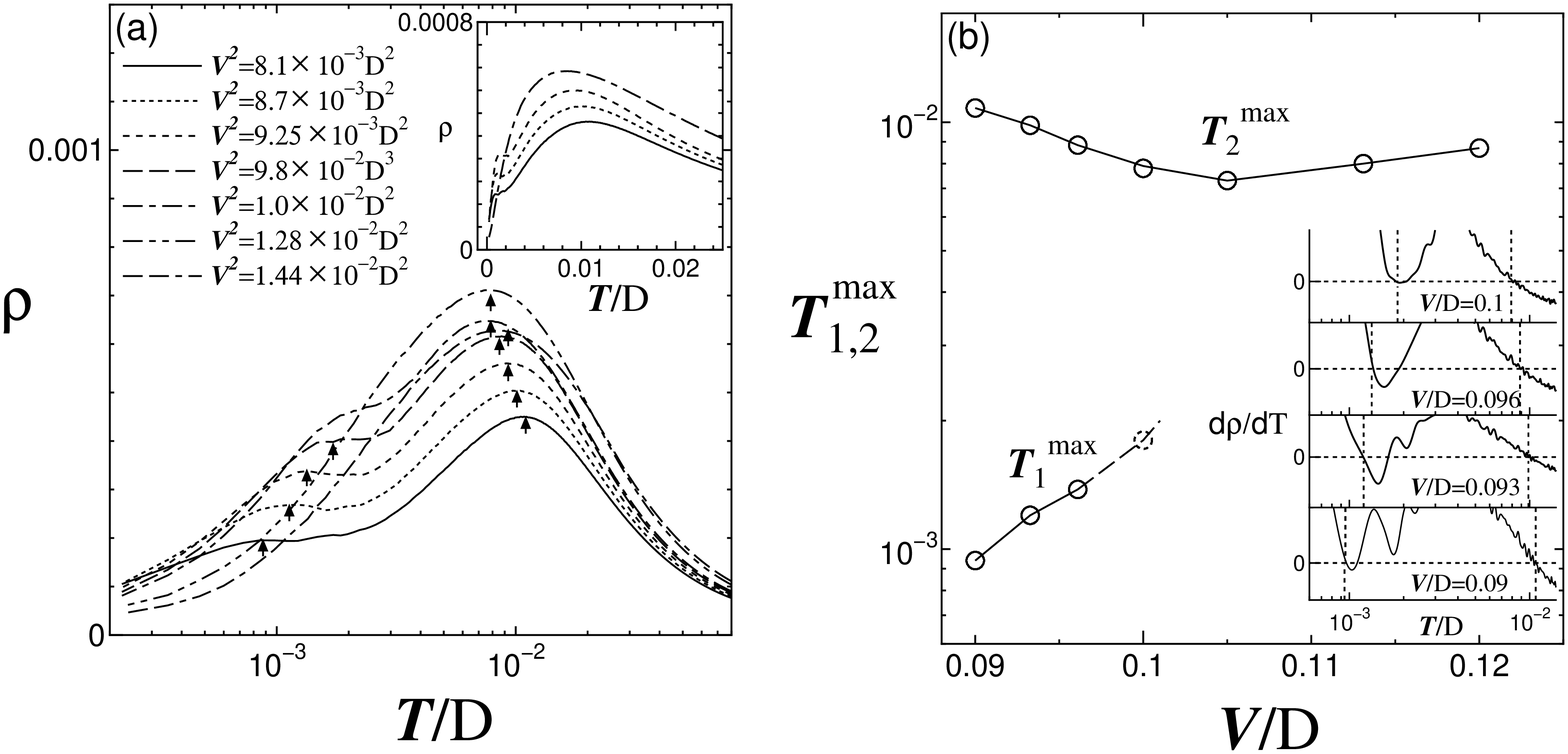}
\end{center}
\caption{(a) Temperature dependence of resistivity for a series of hybridization.
Arrows indicate the temperatures $T_1^{\textrm{max}}$ and $T_2^{\textrm{max}}$
where $-{\textrm{d} \rho}/{\textrm{d} T}$ is zero.
(b) $T^{\textrm{max}}_\textrm{1,2}$ vs $V$ in the case of the
CEF splitting
$\Delta_1=0.02D$, $\Delta_2=0.04D$.}
\label{fig20}
\end{figure}
In this section, we discuss the effect of pressure on the double-peak structure in the $T$-dependence of $\rho(T)$.
In heavy fermion compounds such as $\textrm{Ce}\textrm{Cu}_2\textrm{Si}_2$\cite{Holmes}, the double peaks
merge into a single peak with increasing pressure. We discuss this problem on the basis of the present model 
(infinite-$U$ $J$=5/2 generalized PAM under a tetragonal symmetry) in which the effect of pressure is parameterized
as that of the hybridization $V$ between conduction electrons and f-electrons.
We adopt the following parameters; $D=1$, $n \equiv n_c+n_f=1.4$, $\varepsilon^{(0)}_f=-0.7D$,
and we set the CEF parameters as $\Delta_1=0.02D$ and $\Delta_2=0.04D$ putting the case of $\textrm{Ce}\textrm{Cu}_2\textrm{Si}_2$\cite{Steg} in mind.
We illustrate the results of the calculation for a series of hybridization parameters $V$ in Fig. \ref{fig20}. 
For $V^2 = 8.71\times 10^{-3}D^2\sim 1.0\times 10^{-2}D^2$, the resistivity exhibits a double-peak structure, 
for $V^2 \gtrsim 1.28\times10^{-2}D^2$ the double-peak structure
fades away, and a single-peak structure is obtained.
Namely, we find the double peaks of the resistivity merge into a single peak when
$V$ is increased gradually, which is consistent with experiments under various pressures. 

This result is summarized as follows. The Kondo temperature, which is related to the peak at lower temperatures,
increases with increasing $V$ (See Fig. \ref{fig8}(a)), so that
one peak at a lower temperature shifts to a higher temperature monotonously with increasing $V$.
On the other hand, in Fig. \ref{fig20}(b), another peak at a higher temperature shows only a slight shift to a lower temperature
with increasing $V$. After that, the merged peak, which formed at $V^2\gtrsim 1.28\times10^{-2}D^2$, shifts
further to much higher temperatures with increasing $V$.
This behavior describes the experimental results fairly well.
For a hybridization  $V^2=7.2\times10^{-3}D^2$ 
smaller than $V^2=8.1\times10^{-3}D^2$, the double-peak structure of the resistivity also becomes invisible.
The parameter space in which $\rho(T)$ exhibits the double-peak structure is limited, and 
is dependent on 
the energy scale of $\Delta_1$ and $\Delta_2$ etc.
%
%
%
\section{Conclusions and Discussions}
We have studied the behavior of the electrical resistivity $\rho(T)$ under pressure of $\textrm{Ce}$-based 
heavy fermion systems in the framework of a generalized periodic Anderson model (generalized PAM)
in a manifold of $J=5/2$ under a tetragonal configuration. 
We applied the $1/N$-expansion method to the generalized PAM, and presented numerical results within an accuracy
of $(1/N)^0$ for the calculation at finite temperatures as well as $T=0$. 
Furthermore, by using the present extension of the $(1/N)$-expansion method, 
we investigated the effect of pressure on the resistivity $\rho(T)$ by parameterizing 
the effect of the pressure as a variation of the hybridization
between the conduction electrons and f-electrons.
As a result, we reproduced the behavior that the double-peak structure in $\rho(T)$ is shown to merge into 
a single peak with increasing pressure, which is in good agreement with experiments for numerous heavy fermion compounds.

Here, we point out some remaining problems and perspectives. The present model may be too simplified for quantitative study 
in the sense that we have neglected the ${\bf k}$-dependence of the self-energy as well as the hybridization $V_{{\bf k}\Gamma \sigma}$. 
We may be able to discuss the {\bf k}-dependence of $V_{{\bf k}\Gamma \sigma}$ qualitatively 
by parameterizing the weight of the hybridization parameter $V$ for each level $\Gamma$ without setting $V$ constant.
On the other hand, the ${\bf k}$-dependence of the self-energy is treated by taking into account inter-site correlation which is observed in
the higher order terms of $O(1/N)$. By studying the problem up to $O(1/N)^1$, we can discuss the 
inter-site correlation such as the RKKY interaction as detailed in ref. \citen{kasu}.
Fundamental problems
such as the competition between the RKKY interaction and the Kondo effect 
around quantum critical point in the lattice system remain to be studied further.
%
%
%
\section*{Acknowledgement}
We would like to thank Yoshiaki $\bar{\textrm{O}}$no for valuable discussions and comments.
This work is supported by a Grant-in-Aid for Scientific Research (No.16340103) and 21st Century
COE Program (G18) from the Japan Society for the Promotion of Science.
\section*{Appendix A: Calculation of $G_{f}^{i}(\omega)_{\textrm{(b)}}$}
In this appendix, we discuss the structure of $G_{f}^{i}(\omega)_\textrm{{(b)}}$ in detail.
At $T=0$, the analytical expression of eq. (3.29) is derived, while at a finite temperature $T\gtrsim E_0$
an additional term becomes important.
From the diagrammatic representation in Fig. \ref{fig11}(b),
we write $G_{f}^{i}(\textrm{i}\omega_n)_{\textrm{(b)}}$ as follows:
%
%
%
%
\begin{align}
G_{f}^{i}(\textrm{i}\omega_n)_{\textrm{(b)}}=&\frac{1}{N_L}\sum_{\Gamma\pm}\sum_{\sigma}\sum_{\bf{k}_\sigma}T\sum_{\omega_n'}T\sum_{\nu_n}|V_{\Gamma\sigma}|^2
G_{{\bf k}\sigma}(-\textrm{i}\omega_n')B_i(\textrm{i}\nu_n)\nonumber\\
&\times F^{2}_{i\Gamma}(\textrm{i}\omega_n+\textrm{i}\nu_n)B_i(\textrm{i}\omega_n+\textrm{i}\nu_n+\textrm{i}\omega_n')/\langle \hat{Q} \rangle_{\lambda}. \tag{A.1}
\end{align}
Here, we take the summation with respect to the Matsubara frequency,
%
%
\begin{align}
G_{f}^{i}(\textrm{i}\omega_n)_{\textrm{(b)}}=&-\frac{1}{N_L}\sum_{\Gamma\pm}\sum_{\sigma}\sum_{\bf{k}_\sigma}|V_{\Gamma\sigma}|^2\int \textrm{d}\varepsilon \ f(\varepsilon)
\Bigl[ \frac{-1}{\pi}\textrm{Im}G_{{\bf k}\sigma}(-\varepsilon+\textrm{i}0_+) \Bigr]\nonumber\\
&\times \frac{1}{2\pi \textrm{i}}\int_{C} \textrm{d}\varepsilon' n(\varepsilon')B_i(\varepsilon')F_{i\Gamma}^2(\textrm{i}\omega_n+\varepsilon')B_i(\textrm{i}\omega_n+\varepsilon+\varepsilon')
/\langle \hat{Q}_i \rangle_{\lambda}, \tag{A.2}
\end{align}
where $n(\varepsilon)$ is the Bose distribution function.
Here, we also use the notation: $\bar{B}(\omega)=B_i(\omega+\lambda_i+\varepsilon_f)$ and 
$\langle \hat{Q}_i \rangle_{\lambda}=\textrm{e}^{-\beta(\lambda_i+\varepsilon_f-E_0)} \langle \hat{Q} \rangle$.
Then, the above equation 
can be rewritten as
%
%
\begin{align}
G_{f}^{i}(\textrm{i}\omega_n)_{\textrm{(b)}}=&-\frac{1}{N_L}\sum_{\Gamma\pm}\sum_{\sigma}\sum_{\bf{k}_\sigma}|V_{\Gamma\sigma}|^2\int \textrm{d}\varepsilon\  f(\varepsilon) 
\Bigl[ \frac{-1}{\pi}\textrm{Im}G_{{\bf k}\sigma}(-\varepsilon+\textrm{i}0_+) \Bigr]\nonumber\\
&\times\frac{1}{2\pi \textrm{i}}\int_{C} \textrm{d}\varepsilon' \textrm{e}^{-\beta(E_0+\varepsilon')}\bar{B}(\varepsilon')F_{i\Gamma}^2(\textrm{i}\omega_n+\varepsilon')\bar{B}
(\textrm{i}\omega_n+\varepsilon+\varepsilon')
/\langle \hat{Q}\rangle,\nonumber\\
=&-\frac{1}{N_L}\sum_{\Gamma\pm}\sum_{\sigma}\sum_{\bf{k}_\sigma}|V_{\Gamma\sigma}|^2\int \textrm{d}\varepsilon\ f(\varepsilon) \Bigl[ \frac{-1}{\pi}
\textrm{Im}G_{{\bf k}\sigma}(-\varepsilon+\textrm{i}0_+) \Bigr]\nonumber\\
&\times\frac{1}{2\pi \textrm{i}}\int_{C} \textrm{d}\varepsilon'\ \textrm{e}^{-\beta(E_0+\varepsilon')}\bar{B}(\varepsilon')F_{i\Gamma}^2(\textrm{i}\omega_n+\varepsilon')\bar{B}(\textrm{i}\omega_n+\varepsilon+\varepsilon')
/\langle \hat{Q} \rangle,\nonumber\\
=&\frac{1}{N_L}\sum_{\Gamma\pm}\sum_{\sigma}\sum_{\bf{k}_\sigma}|V_{\Gamma\sigma}|^2\int \textrm{d}\varepsilon \ f(\varepsilon) \Bigl[ \frac{-1}{\pi}
\textrm{Im}G_{{\bf k}\sigma}(-\varepsilon+\textrm{i}0_+) \Bigr]/\langle \hat{Q}\rangle\nonumber\\
&\times \Biggl[
\int \textrm{d}\varepsilon' \ \textrm{e}^{-\beta(E_0+\varepsilon')}\Bigl[ \frac{-1}{\pi}
\textrm{Im}\bar{B}(\varepsilon'+\textrm{i}0_+)\Bigr]\bar B(\textrm{i}\omega_n+\varepsilon+\varepsilon') \frac{1}{(\textrm{i}\omega_n+\varepsilon'-E_{i\Gamma}+\varepsilon_f)^2}\nonumber\\
&+\int \textrm{d}\varepsilon' \ \textrm{e}^{-\beta(E_0-\varepsilon+\varepsilon')}\Bigl[ \frac{-1}{\pi}
\textrm{Im}\bar{B}(\varepsilon'+\textrm{i}0_+)\Bigr]
\bar B(-\textrm{i}\omega_n-\varepsilon+\varepsilon') \frac{-1}{(\varepsilon'-\varepsilon-E_{i\Gamma}+\varepsilon_f)^2}\nonumber\\
& +\frac{\textrm{d}}{\textrm{d}\varepsilon'}\Bigl[\textrm{e}^{-\beta(E_0+\varepsilon')}\bar{B}(\varepsilon')\bar{B}(\varepsilon+\varepsilon'+\textrm{i}\omega_n)   \Bigr]\Big|_{\varepsilon'=-\textrm{i}\omega_n+E_{i\Gamma}-\varepsilon_f}
\Biggr]. \tag{A.3}
\end{align}
After performing the analytic continuation $\textrm{i}\omega_n \rightarrow \omega_{+}\equiv \omega+\textrm{i}0_+$,
and changing the variable $\varepsilon \rightarrow -\varepsilon$ in the second term,
we obtain eq. (3.29).
At $T=0$, the third term, which is the contribution from the double pole of the pseudo-fermion,
vanishes because of a factor $\exp(-\beta E_0)$. 

At a finite temperature ($T \gtrsim E_0$), however, the third term in eq. (A.3) is non-vanishing.
The contribution from $G_{f}^{i}(\omega)_{\textrm{(b)}}$ in Fig. \ref{fig11}(b), 
consists of both the pole of the slave-boson,
${G_{f}^{i}}^{(bp)}(\omega)_{\textrm{(b)}}$, and the double poles of the pseudo-fermion, 
${G_{f}^{i}}^{(dp)}(\omega)_{\textrm{(b)}}$.
Here, we define ${\rho_f^{i}}(\omega)_{\textrm{(b)}}=-1/\pi\textrm{Im}{G_{f}^{i}}(\omega)_{\textrm{(b)}}$,
${\rho_f^{i}}^{(bp)}(\omega)_{\textrm{(b)}}=-1/\pi \textrm{Im}{G_{f}^{i}}^{(bp)}(\omega)_{\textrm{(b)}}$,
and ${\rho_f^{i}}^{(dp)}(\omega)_{\textrm{(b)}}=-1/\pi \textrm{Im}{G_{f}^{i}}^{(dp)}(\omega)_{\textrm{(b)}}$.
Then,
%
%
\begin{align}
&\rho^{i}_f(\omega)_{\textrm{(b)}}={\rho_f^{i}}^{(bp)}(\omega)_{\textrm{(b)}}
+{\rho_f^{i}}^{(dp)}(\omega)_{\textrm{(b)}}, \tag{A.4}
\end{align}
%
%
%
%
%
\begin{figure}
\begin{center}
\includegraphics[scale=0.4]{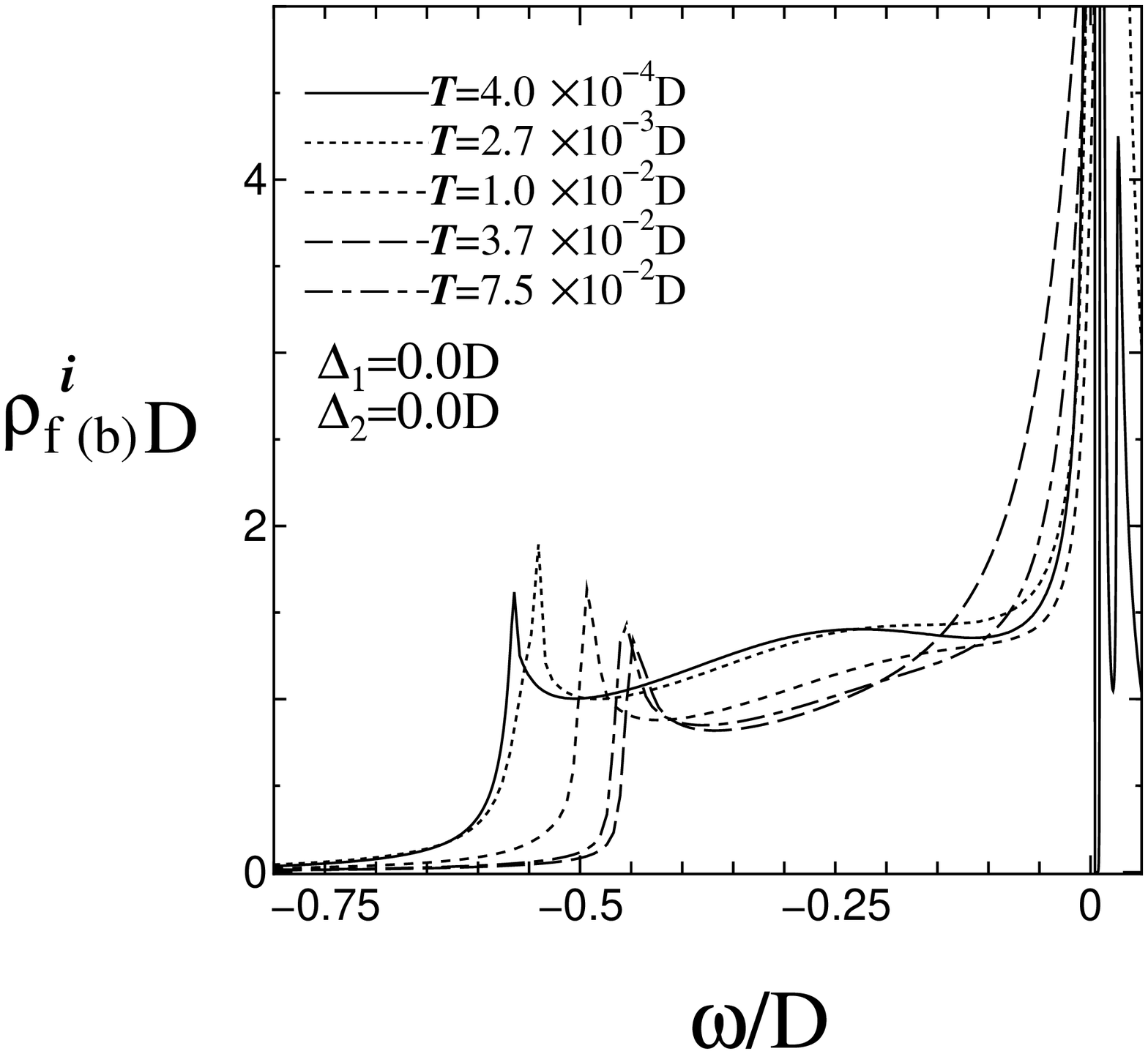}
\end{center}
\caption{${\rho^{i}_f}(\omega)_{\textrm{(b)}}$ vs $\omega$ for a series of temperatures. }
\label{fig21}
\end{figure}
where
%
%
\begin{align}
{\rho_f^{i}}^{(bp)}(\omega)_{\textrm{(b)}}=&\sum_{\Gamma\pm}\sum_{\sigma}|V_{\Gamma\sigma}|^2\int \textrm{d} \varepsilon \ f(\varepsilon) 
\int \textrm{d} \varepsilon'
\textrm{e}^{-\beta(E_0+\varepsilon')} \frac{-1}{\pi}\textrm{Im}\bar{B}(\varepsilon '+\textrm{i}0_+)/\langle \hat{Q} \rangle\nonumber\\
&\times \Bigl[ \rho_\sigma(-\varepsilon)\cdot \frac{-1}{\pi}\textrm{Im}\bar{B}(\omega+\varepsilon+\varepsilon'+\textrm{i}0_+)
\cdot \frac{1}{(\omega+\varepsilon'-E_{i\Gamma}+\varepsilon_f)^2}\nonumber\\
&+\rho_\sigma(\varepsilon)\cdot \frac{-1}{\pi}\textrm{Im}\bar{B}(-\omega+\varepsilon+\varepsilon'+\textrm{i}0_+) \cdot \frac{1}{(\varepsilon+\varepsilon'-E_{i\Gamma}+\varepsilon_f)^2} \Bigr],\tag{A.5}
\end{align}
and
%
%
\begin{align}
{\rho_f^{i}}^{(dp)}(\omega)_{\textrm{(b)}}=&\sum_{\Gamma\pm}\sum_{\sigma}|V_{\Gamma\sigma}|^2
\int \textrm{d} \varepsilon \ f(\varepsilon) \rho_\sigma(-\varepsilon)\cdot (\frac{-1}{\pi})\cdot \textrm{e}^{-\beta (E_0+E_{i \Gamma}-\varepsilon_f)} /\langle \hat{Q} \rangle\nonumber\\
&\times \Bigl[  -\frac{1}{T} \textrm{Im}{\bar B}(-\omega+E_{i\Gamma}-\varepsilon_f+\textrm{i}0_+)
{\bar B(\varepsilon}+E_{i\Gamma}-\varepsilon_f)\nonumber\\
&+\bar{B}(\varepsilon+E_{i\Gamma}-\varepsilon_f)\textrm{Im}\frac{\textrm{d} \bar{B}(\varepsilon'+\textrm{i}0_+)}
{\textrm{d}\varepsilon'}\Bigr|_{\varepsilon'=-\omega+E_{i\Gamma}-\varepsilon_f}\nonumber\\
&+\textrm{Im}\bar{B}(-\omega+E_{i\Gamma}-\varepsilon_f+\textrm{i}0_+)
\frac{\textrm{d}\bar{B}(\varepsilon')}{\textrm{d}\varepsilon'}\Bigr|_{\varepsilon'=\varepsilon+E_{i\Gamma}-\varepsilon_f} \Bigr]. \tag{A.6}
\end{align}
In Fig. \ref{fig21}, we show the temperature dependence of ${\rho^{i}_f}(\omega)_{\textrm{(b)}}$ as calculated numerically.
%
%
\section*{Appendix B:   Kondo temperature in impurity case}
In this appendix, we derive the relation between the Kondo temperature and the CEF-splitting
in the impurity Anderson model, eq. (3.36).
$E_0$ is defined as the binding energy of the slave-boson. At $T=0$, $E_0$
corresponds to the Kondo temperature $T_K$
defined in an impurity version of the Anderson model. 
In the impurity problem, the conduction electron is not renormalized, 
so  we can put  $E_0 \rightarrow T_K$,
$\alpha_{{\bf k}\sigma}^{j}\rightarrow \varepsilon_{{\bf k}\sigma}$ and $A_{\bf k}^{j}\rightarrow 1$, 
in eq. (3.12):
%
%
\begin{align}
T_K-\varepsilon_f=& \frac{1}{N_L}\sum_{\Gamma\pm}\sum_{\sigma}\sum_{{\bf k}_\sigma}\frac{f(\varepsilon_{{\bf k}\sigma})|V_{\Gamma\sigma}|^2}
{T_K+E_{i\Gamma}-\varepsilon_f-\varepsilon_{{\bf k}\sigma}},\nonumber\\
=&\frac{V^2}{N_L}\sum_{\sigma}\sum_{{\bf k}_\sigma}\frac{f(\varepsilon_{{\bf k}\sigma})}{T_K-\varepsilon_{{\bf k}\sigma}}
+\frac{V^2}{N_L}\sum_{\sigma}\sum_{{\bf k}_\sigma}\frac{f(\varepsilon_{{\bf k}\sigma})}{T_K+\Delta_1-\varepsilon_{{\bf k}\sigma}},\nonumber\\
&+\frac{V^2}{N_L}\sum_{\sigma}\sum_{{\bf k}_\sigma}\frac{f(\varepsilon_{{\bf k}\sigma})}{T_K+\Delta_2-\varepsilon_{{\bf k}\sigma}}, \tag{B.1}
\end{align}
where we have used the following approximation: $\sum_{\pm}|V_{\Gamma\sigma}|^2=\sum_{\pm}|V_{\Gamma\bar{\sigma}}|^2=V^2$, where
$\sum_{\pm}$ means the summation over the Kramers doublets.
Here, we assume $T_K \ll \Delta_1\sim\Delta_2 \ll |\varepsilon_f| < |D|$, which is
a reasonable assumption in an actual system,
and $ ({1}/{N_L}){\sum_{{\bf k}}}$
is replaced by the integral $\rho_0\int_{-D}^{D} \textrm{d}\varepsilon $, 
$\rho_0$ being the DOS of the conduction electron
per spin at Fermi level. Then, eq. (B.1) is reduced to
%
%
\begin{align}
-\varepsilon_f  \simeq \ & 2V^2 \rho_0 \int_{-D}^{D} \textrm{d}\varepsilon\  f(\varepsilon)\frac{1}{T_K-\varepsilon}\nonumber\\
&+2V^2 \rho_0 \int_{-D}^{D}  \textrm{d}\varepsilon \ f(\varepsilon)\frac{1}{\Delta_1-\varepsilon}+2V^2 \rho_0 \int_{-D}^{D} \textrm{d}\varepsilon \ f(\varepsilon)\frac{1}{\Delta_2-\varepsilon}. \tag{B.2}
\end{align}
On the basis of this equation at $T=0$, the Kondo temperature $T_K$ is evaluated as follows:
$\varepsilon$-integration in (B.2) is performed explicitly, leading to
%
%
\begin{align}
\varepsilon_f=& 2V^2 \rho_0 \log \Bigl|\frac{D-T_K}{T_K} \Bigr|
+2V^2 \rho_0 \log \Bigl|\frac{D-\Delta_1}{\Delta_1} \Bigr|+2V^2 \rho_0 \log \Bigl|\frac{D-\Delta_2}{\Delta_2} \Bigr|\nonumber,\\
\simeq & 2V^2 \rho_0 \log \Bigl|\frac{D}{T_K} \Bigr|
+2V^2 \rho_0 \log \Bigl|\frac{D}{\Delta_1} \Bigr|
+2V^2 \rho_0 \log \Bigl|\frac{D}{\Delta_2} \Bigr|. \tag{B.3}
\end{align}
Then, it is easy to find a solution for $T_{K}$ as
%
%
\begin{align}
T_K
=&D\cdot(\frac{D}{\Delta_1})\cdot(\frac{D}{\Delta_2})\exp\Bigl(
-\frac{|\varepsilon_f|}{2\rho_0V^2} \Bigr).\tag{B.4}
\end{align}
Similarly, we can calculate the Kondo temperature $T_K^{(0)}$ in the absence of CEF splitting.
By putting  $\Delta_1=\Delta_2=0$ in eq. (B.1), we obtain
%
%
\begin{align}
T_K^{(0)}-\varepsilon_f=&3\frac{V^2}{N_L}\sum_{\sigma}\sum_{{\bf k}_\sigma}\frac{f(\varepsilon_{{\bf k}\sigma})}
{T_K^{(0)}-\varepsilon_{{\bf k}\sigma}}. \tag{B.5}
\end{align}
Hence,
%
%
\begin{align}
T_K^{(0)}=D\exp\Bigl(-\frac{|\varepsilon_f|}{6\rho_0V^2} \Bigr). \tag{B.6}
\end{align}
From the relations between $T_K$ and $T_K^{(0)}$, we can derive equation (3.36):
%
%
\begin{align}
\Bigl[ T_K^{(0)} \Bigr]^3= \Delta_1\cdot\Delta_2\cdot T_K. \tag{B.7}
\end{align}
In the leading order, the result with use of $1/N$-expansion method in case of the impurity problem,
is consistent with the previous impurity study\cite{yama,han}.
%
%
%
\section*{Appendix C: The algorithm for self-consistent $1/N$-expansion}
In this appendix, we explain the algorithm of self-consistent $1/N$-expansion over the entire temperature range.
The whole procedure is summarized as follows:
\\
1) Self-energy of conduction electron

First, we choose the self-energy $\Sigma_{{\bf k}\sigma}(\omega)$
of the conduction electron as a trial function.
If we start from the low-temperature region, we can choose $\Sigma_{{\bf k}\sigma}(\omega)$ 
derived from low-temperature approximation, eq. (3.6), i.e.,
%
%
\begin{align}
\Sigma_{{\bf k}\sigma}(\omega)=\sum_{\Gamma\pm}\frac{a|V_{{\bf k}\Gamma \sigma}|^2}
{\omega-E_0-E_{i\Gamma}+\varepsilon_f},\tag{C.1}
\end{align}
where $E_0$ and $a$ is obtained by the coupled self-consistent equations at $T=0$,
eqs. (3.12) $\sim$ (3.14).
We restrict the range of the energy, from $-\omega_\textrm{max}$ to $\omega_\textrm{max}$. 
In an actual numerical calculation, $\omega$ is defined as a discrete value $\omega_{j}$ 
($j=-{j}_\textrm{max},\ -{j}_\textrm{max}+1,\ \cdots\  ,{j}_\textrm{max}-1,\ {j}_\textrm{max}$), 
where $j_\textrm{max}$ is an integer of maximum cut number of the energy.
For example, we define $\omega_j$ as follows:
%
%
\begin{align}
\omega_j=\frac{j^2}{j^2_\textrm{max}}\omega_\textrm{max}.\tag{C.2}
\end{align}
We use $\omega_\textrm{max}= 2D$ and $j_\textrm{max}=700$ in the present paper.
\\
2) Green function of the conduction electron

The Green function of the conduction electron is obtained from the above self-energy as
%
%
\begin{align}
G_{{\bf k}\sigma}(\omega)=\frac{1}{\omega-\varepsilon_{{\bf k}\sigma}+\mu-\Sigma_{{\bf k}\sigma}(\omega)},\tag{C.3}
\end{align}
where $\mu$ is the chemical potential.
Then, $(\frac{1}{N_L})\sum_{{\bf k}_\sigma}G_{{\bf k}\sigma}(\omega)$ is performed as follows,
%
%
\begin{align}
\frac{1}{N_L}\sum_{{\bf k}_\sigma}G_{{\bf k}\sigma}(\omega)
&=\rho_0\int^{D}_{-D}
\textrm{d}\varepsilon \frac{1}{\omega-\varepsilon+\mu-\Sigma_{{\bf k}\sigma}(\omega)},\nonumber\\
&=\frac{1}{2D}\log  \frac{\omega+D+\mu-\Sigma_{{\bf k}\sigma}(\omega)}
{\omega-D+\mu-\Sigma_{{\bf k}\sigma}(\omega)},\tag{C.4}
\end{align}
where $\rho_0\equiv1/2D$ as in eq. (3.16), and $\mu\equiv 0$ in our numerical calculation.\\
3) Self-energy of slave-boson

The imaginary part of the self-energy $\textrm{Im}\bar{\Pi}(\omega)$, eq. (2.51),
of the slave-boson is calculated by using $G_{{\bf k}\sigma}$, eq. (C.3),
as
%
%
\begin{align}
\textrm{Im}\bar{\Pi}(\omega)=\sum_{\Gamma\pm}\sum_{\sigma}\sum_{{\bf k}_\sigma}
\frac{|V_{{\bf k}\Gamma\sigma}|^2}{N_L}f(-\omega+E_{i\Gamma}-\varepsilon_f)\textrm{Im}G_{{\bf k}\sigma}
(-\omega+E_{i\Gamma}-\varepsilon_f).\tag{C.5}
\end{align}
$\textrm{Re}\bar{\Pi}(\omega)$ is calculated by the Kramers-Kr\"{o}nig relation.
Hence, we can obtain $\bar{\Pi}
(\omega)=\textrm{Re}\bar{\Pi}(\omega)+\textrm{i}\cdot\textrm{Im}\bar{\Pi}(\omega)$\\
4) Green function of slave-boson

The Green function $\bar{B}(\omega)$ of the slave-boson is given as
%
%
\begin{align}
\bar{B}(\omega)=\frac{1}{\omega+\varepsilon_f-\bar{\Pi}(\omega)}.\tag{C.6}
\end{align}
5) Determination of $E_0(T)$

The binding energy of the slave-boson $E_0(T)$ defined as the real part of the pole 
of $\bar{B}(\omega)$, eq. (C.6). However, in actual calculation, we approximate it as the
value which satisfies the following equation:
%
%
\begin{align}
\varepsilon_f-E_0(T)=\textrm{Re}\bar{\Pi}(-E_0(T)).\tag{C.7}
\end{align}
This equation is an approximation of eq. (2.54).\\
%
%
%
%
%
%
%
\begin{figure}[t]
\begin{center}
\includegraphics[scale=0.35]{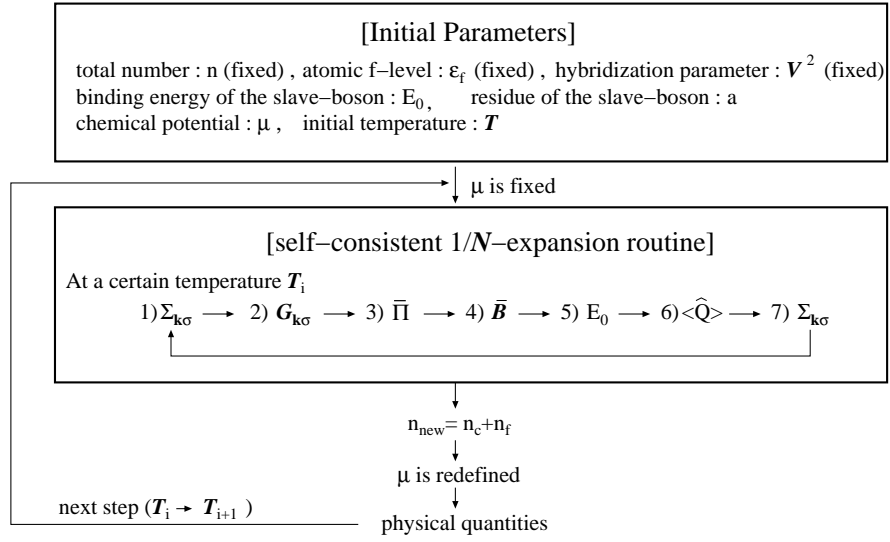}
\end{center}
\caption{
An example algorithm for self-consistent $1/N$-expansion.}
\label{fig22}
\end{figure}
6) Calculation of $\langle \hat{Q}_i \rangle_{\lambda}$

We have defined $\langle \hat{Q}_i \rangle_{\lambda}=\textrm{e}^{-\beta(\lambda_i+\varepsilon_f-E_0(T))}
\langle \hat{Q} \rangle$, where $E_0(T)$ is obtained by procedure 5). 
We calculate $\langle \hat{Q} \rangle$ from eqs. (2.39), (2.48), and (2.49) by using the Green function of the conduction electron
in procedure 2), 
the slave-boson in procedure 4), and the pseudo-fermion.\\
7) Self-energy of conduction electron

Finally, through the procedures 2) $\sim$ 6), 
we can calculate $\Sigma_{{\bf k}\sigma}$ again by eq. (2.43) as
%
%
\begin{align}
\textrm{Im}\Sigma_{{\bf k}\sigma}(\omega)=\sum_{\Gamma\pm}|V_{{\bf k}\Gamma\sigma}|^2
\Bigr( 1+\textrm{e}^{\beta \omega}\Bigl)
 \textrm{e}^{-\beta(E_0(T)+E_{i\Gamma}-\varepsilon_f)}\cdot
\textrm{Im}\bar{B}(-\omega+E_{i\Gamma}-\varepsilon_f). \tag{C.8}
\end{align}
$\textrm{Re}\Sigma_{{\bf k}\sigma}(\omega)$ is derived from the 
Kramers-Kr\"{o}nig relation.
As a result, we can obtain the newly defined
$\Sigma_{{\bf k}\sigma}(\omega)\equiv \textrm{Re}\Sigma_{{\bf k}\sigma}(\omega)+\textrm{i}\cdot\textrm{Im}\Sigma_{{\bf k}\sigma}(\omega)$.

This series of procedures, 1)$\sim$7), forms a self-consistent cycle.
We show the concept chart of this cycle in Fig. \ref{fig22}.
In our calculation, this procedure was performed with fixed $n$, the total number of electron per site.
In order to check the self-consistency, we define the following amount,
%
%
\begin{align}
\delta{\textrm{Im}\bar{\Pi}}
&=\frac{\displaystyle{\sum^{j_{\textrm{max}}}_{j=-j_{\textrm{max}}}}|\textrm{Im}\bar{\Pi}_{\textrm{new}}(\omega_j)
-\textrm{Im}\bar{\Pi}_{\textrm{old}}(\omega_j)|
}{2j_{\textrm{max}}+1}
.\tag{C.9}
\end{align}
$\textrm{Im}\bar{\Pi}_{\textrm{new}}(\omega)$ is a new $\textrm{Im}\bar{\Pi}(\omega)$ 
obtained from one cycle by using $\textrm{Im}\bar{\Pi}_{\textrm{old}}(\omega)$.
This self-consistent cycle is repeated until $\delta{\textrm{Im}\bar{\Pi}}(\omega)$ becomes less than $10^{-4}D\sim 10^{-5}D$.
By performing this procedure for each temperature, we can calculate the temperature 
dependence of physical quantities such as the electrical resistivity.
%
%
%
%

\end{document}